\documentclass[journal]{IEEEtran}
\usepackage{stfloats}

\usepackage{enumerate}
\usepackage{algorithm,algpseudocode}
\usepackage{setspace,amsmath,latexsym,cite,amssymb,epsfig,amsfonts}
\usepackage{url,cite}
\usepackage{graphicx}
\usepackage{psfrag}
\usepackage{footmisc}
\usepackage{multirow}
\usepackage{color}
\usepackage{multicol}
\usepackage{mathtools}
\usepackage{subcaption}
\usepackage{graphicx}
\usepackage{amssymb}
\usepackage{amsmath}
\usepackage{epstopdf}
\usepackage{geometry}
\usepackage{float}
\usepackage{balance}
\usepackage{makecell}
\usepackage{changepage}

\algnewcommand\algorithmicforeach{\textbf{for}}
\algdef{S}[FOR]{For}[1]{\algorithmicforeach\ #1\ \algorithmicdo}

\twocolumn

\makeatother

        \makeatletter
        \def\fps@eqnfloat{!t}
        \def\ftype@eqnfloat{4}
        
        \newenvironment{eqnfloat*}
               {\@dblfloat{eqnfloat}}
               {\end@dblfloat}
        \makeatother
\allowdisplaybreaks[4]

\title{Hybrid Bit and Semantic Communications for UAV-Enabled Wireless Power Transfer Networks:\\A Decision-Assisted Deep Reinforcement\\Learning Approach}
\author{Jingfu Li, \IEEEmembership{Member, IEEE}, Jingjing Cui, \IEEEmembership{Senior Member, IEEE}, Chong Huang, \IEEEmembership{Member, IEEE}, \\Jing Zhu, \IEEEmembership{Member, IEEE}, Zheng Chu, \IEEEmembership{Member, IEEE}, Mingzhe Chen, \IEEEmembership{Senior Member, IEEE}, \\Pei Xiao, \IEEEmembership{Senior Member, IEEE}, Rahim Tafazolli, \IEEEmembership{Fellow, IEEE}
\thanks{The work presented in this article was supported by the U.K. Engineering and Physical Sciences Research Council under Grant EP/X013162/1. }
\thanks{J. Li and J. Cui are with School of Information Science and Technology, Southwest Jiaotong University, Chengdu, Sichuan, China (e-mail: jingfuli@swjtu.edu.cn, jingjing.cui@swjtu.edu.cn).}
\thanks{C. Huang, P. Xiao and R. Tafazolli are with 5GIC \& 6GIC, Institute for Communication Systems (ICS), University of Surrey, Guildford, GU2 7XH, United Kingdom (e-mail: chong.huang@surrey.ac.uk, p.xiao@surrey.ac.uk, r.tafazolli@surrey.ac.uk).}
\thanks{Jing Zhu is with the School of Flexible Electronics (SoFE) \& State Key Laboratory of Optoelectronic Materials and Technologies (OEMT), Sun Yat-sen University, Shenzhen, Guangdong 518107, China (e-mail: zhuj229@mail.sysu.edu.cn).}
\thanks{Z. Chu is with the Department of Electrical and Electronic Engineering, University of Nottingham Ningbo China, Ningbo 315100, China (Email: andrew.chuzheng7@gmail.com).}
\thanks{M. Chen is with the Department of Electrical and Computer Engineering, University of Miami, FL, 33146, USA (Email: mingzhe.chen@miami.edu).}
}

\begin{document}
\captionsetup[figure]{name={Fig.},labelsep=period}

\begin{singlespace}
\maketitle

\end{singlespace}

\thispagestyle{empty}
\begin{abstract}
Semantic communications which can significantly reduce spectrum consumption in wireless networks, have recently become a popular research area. When combined with wireless power transfer (WPT), semantic communications can help achieve high spectral efficiency for energy-limited devices in wireless communications. In energy-constrained and link budget-limited scenarios such as UAV networks, the integration of semantic communications and WPT enables highly energy-efficient transmission mechanisms. In this paper, we investigate semantic communications in UAV-enabled WPT networks. To achieve adaptability to varying signal-to-noise ratio (SNR) and task requirements, we introduce a multi-layer hybrid bit and semantic communication framework. We adopt a semantic communication efficiency metric and aim to maximize it by jointly optimizing UAV trajectory, energy harvesting base station (EHBS) selection, user association, semantic mode selection, and energy harvesting time allocation. To address this complex long-term optimization problem, we introduce the distributional soft actor-critic (DSAC) algorithm and introduce a decision assistant to further enhance the convergence performance of DSAC. Simulation results validate the effectiveness of the proposed method and framework and demonstrate that our algorithm can achieve superior long-term optimization performance in dynamic network environments.
\end{abstract}


\section{Introduction}
In this section, we introduce the background and related works of semantic communications and UAV-enabled wireless power transfer (WPT) networks. We then summarize the main contributions of this paper.

\subsection{Background}
With the development of sixth-generation (6G), semantic communications have emerged as a key technique that focuses on semantic information of data instead of bitwise fidelity, thereby enhancing spectrum efficiency in wireless communications \cite{lu2024tutorial}. Semantic communications introduce the semantic encoder to extract semantic features from the raw data, and the semantic decoder to reconstruct the raw data based on the semantic information from the encoder. Therefore, semantic communications can significantly reduce the bandwidth requirement for wireless transmissions in 6G. Recently, semantic communications have been applied in various fields, e.g., DeepSC for text transmissions \cite{xie2021tsp}, DeepJSCC for image transmissions \cite{bourtsoulatze2019deepjscc}, and adaptive semantic model under signal-to-noise ratio (SNR) variations \cite{gunduz2023jsac}.

On the other hand, wireless power transfer (WPT) has become an attractive area in current wireless communication research because it can help realize energy harvesting via radio frequency (RF) \cite{bi2015wpcn}. WPT will bring new opportunities to future communication networks by enabling active, controllable, and truly mobile energy delivery without the need for power cables or large-capacity batteries, thereby facilitating the development of smaller and lighter mobile devices. WPT has already found wide applications in areas such as low-power sensing, low-power computing, and wireless charging systems, providing a foundation for sustainable and energy-efficient wireless communications. Moreover, the transmit power can be further increased with the development of antenna arrays \cite{11173626}, which is highly beneficial for wireless power transfer.

In recent years, unmanned aerial vehicles (UAVs) have attracted increasing attention in wireless communications due to their high flexibility and mobility. UAVs can provide flexible coverage, but maintaining high-rate wireless transmissions is a significant challenge due to their limited battery energy \cite{zeng2016uav}. WPT has become a significant technology in UAV energy consumption research \cite{9468714}. This technology is particularly meaningful for UAVs with limited energy capacity. In wireless power transfer (WPT)–enabled UAV networks, trajectory planning and resource allocation can significantly improve energy transfer efficiency and transmission rates \cite{yuan2021twc}. However, due to the power limitations and energy loss inherent in WPT, UAVs still cannot perform continuous high-power data transmission even with WPT support. Since semantic communication can save spectrum resources, thereby reducing transmit power or overall transmission time, the integration of semantic communication and WPT can enable UAVs to operate as aerial communication nodes with extended service time and improved quality of service.

\subsection{Related Work}\label{sec:RW}
At present, semantic communication has attracted widespread attention and research in future 6G networks. In \cite{9763856}, semantic information is optimized according to channel conditions, thereby improving the semantic efficiency of text transmission. An end-to-end encoder–decoder based on deep learning was proposed to realize robust semantic information in text transmissions \cite{xie2021tsp}. To transmit images in semantic communications, separate source/channel codes with a convolutional autoencoder were studied in wireless communications to realize the DeepJSCC method \cite{bourtsoulatze2019deepjscc}. Deep learning is introduced to construct a semantic communication framework, enhancing the performance efficiency of multimedia transmission in \cite{9450827}. In \cite{9953076}, deep learning techniques are employed to extract and reconstruct the original image information in semantic communication. To improve the quality of experience (QoE) for users, \cite{10508293} integrates a semantic communication system into an edge computing framework to assist with computation offloading tasks. In \cite{witt2022}, channel state information (CSI)-aware modulation was investigated to enhance the image transmission quality in semantic communications. To study an adaptive rate method in semantic communications, incremental knowledge-based hybrid automatic repeat request (HARQ) was introduced to control the semantic bitrate in dynamic wireless environments \cite{zhou2022adaptive}. In \cite{11006980}, an iterative algorithm is employed in space–air–ground integrated networks (SAGINs) to optimize the UAV’s position, computational capability, semantic compression ratio, task allocation, and other communication resources to reduce the energy consumption of semantic communication. A multi-scale semantic extractor was studied based on Vision Transformer framework to enhance the semantic communication performance in image transmissions \cite{10854360}.

On the other hand, UAV communication has become an important component of communication networks \cite{9768113}. In \cite{10770152}, a channel modeling and prediction method based on UAV trajectory, vibration, and channel sparsity conditions was proposed. To study the three dimensional positioning in UAV networks, UAV trajectory and fluid antenna port were optimized by a deep reinforcement learning method to enhance the average positioning accuracy in \cite{11216397}. In \cite{11123630}, convex optimization techniques were applied to optimize UAV trajectory and beamforming design, thereby improving the secure communication rate of space–air–maritime networks. To suppress interference in space–air–ground networks, interference alignment design and degrees-of-freedom analysis were employed in UAV–reconfigurable intelligent surface (RIS)-assisted communication systems in \cite{10499205}. In \cite{10287142}, UAV trajectory and RIS phase were jointly optimized to enhance the secrecy communication rate in space–air–ground cognitive networks. To address the 3D tracking problem of UAVs, collaborative reinforcement learning is employed in \cite{10483540} to jointly optimize the UAV’s transmit power and trajectory planning, thereby improving the positioning accuracy. In \cite{10680056}, to investigate the mobile user tracking and robust beamforming design, UAV was utilized in SAGINs to enhance the energy efficiency. To study the covert communication performance in UAV networks, flexible RIS was utilized to enhance the transmission efficiency by adjusting RIS phases and incident angle \cite{11271691}. In \cite{11107245}, UAV trajectory and resource allocation was optimized by deep reinforcement learning methods to improve the average throughput in visible light communications.

Currently, energy limitation is a critical challenge in the research area of UAV \cite{11320813}. To address this issue, the integration of wireless power transfer with UAVs has become a popular and increasingly active research direction. In \cite{9769985}, reinforcement learning was employed to investigate the WPT performance in RIS-assisted UAV Internet of Things (IoT) networks, thereby improving the overall network throughput. Convex optimization was utilized to jointly optimize UAV trajectories and RIS phase shifts to minimize energy consumption in WPT-assisted UAV communications in \cite{9708417}. In \cite{10033084}, genetic clustering algorithms and dynamic clustering strategies were applied to jointly optimize UAV control, maximizing UAV energy utilization efficiency. To enhance the throughput of UAV relay networks, UAV trajectory and transmit power were jointly optimized under WPT conditions in \cite{10399860}.

Semantic communications can reduce energy consumption by lowering spectrum resource requirements, making its integration with WPT particularly suitable for future green communication. In \cite{liew2022icassp}, a deep learning–based optimal auction mechanism was proposed to optimize energy allocation strategies and enhance WPT efficiency in IoT networks. To analyze the relationship between semantic communications and energy transfer, the achievable and inverse regions were derived from an information-theoretic perspective to characterize the corresponding energy regions \cite{khalfet2025tcomm}. In \cite{delfani2025lcomm}, the Version Age of Information (VAoI) in satellite communication networks was updated by optimizing semantic-aware strategies, significantly improving energy efficiency. Through joint optimization of communication resources and raw data selection, deep reinforcement learning was used to enhance total data transmission in IoT networks in \cite{sang2025iotj}.

However, the above works have not considered the joint integration of WPT-based energy transfer and semantic communications’ adaptive channel capability, nor have they explored the advantages of hybrid bit and semantic communications. Moreover, the combination of UAVs, WPT, and semantic communications remains an open research area.

\subsection{Motivation and Contributions}
Since semantic communications improve spectrum utilization, its integration with WPT can further enhance wireless transmission efficiency—particularly benefiting energy-constrained UAV networks. Furthermore, a hybrid bit and semantic communication mode can offer more flexibility in transmission performance. Therefore, we investigate a UAV semantic communication system assisted by WPT. In this system, the UAV harvests energy from an energy harvesting base station (EHBS) in each time slot and serves users using an adaptive hybrid bit and semantic communication strategy. Our main contributions are as follows:

\begin{itemize}
\item We propose a WPT-assisted UAV semantic communication system, where the UAV harvests energy from the EHBS based on a non-linear wireless power transfer model in each time slot and uses the harvested energy to transmit data to ground users, and ground users move based on the random waypoint model.

\item We design a hybrid bit and semantic communication framework, which includes bit transmission and multi-level semantic communication modes, enabling adaptation to different transmission conditions. Moreover, we introduce a semantic efficiency metric to evaluate overall transmission performance.

\item We employ a decision-assisted deep reinforcement learning method to jointly optimize UAV 3D trajectory, EHBS selection, user association, and semantic communication mode under random task arrivals, thereby improving semantic efficiency in UAV-enabled WPT systems.

\item Simulation results demonstrate the effectiveness of the proposed algorithm. The hybrid transmission mode adapts well to variations in energy harvesting and dynamic channels, while deep reinforcement learning enhances optimization performance through interaction with the simulation environment, ensuring efficient policy learning in dynamic UAV networks.
\end{itemize}

The rest of this paper is organized as follows: In Section \ref{sec:sm}, the UAV-enabled WPT system model is introduced. The hybrid bit and semantic communication framework, the semantic efficiency metric, and problem formulation are described in \ref{sec:semantic}. In Section \ref{sec:DSAC}, the decision-assisted DRL-based resource allocation algorithm is proposed. Section \ref{sec:sim} presents the simulation results and analysis. Finally, Section \ref{sec:con} concludes this paper.

\begin{table}[t]
 \caption{Main Parameter Notations}
  \centering
  \begin{tabular}{|c|c|}
  \hline
  Number of EHBS & $M$\\
  \hline
  Number of GUs & $N$\\
  \hline
  UAV Trajectory & $q$ \\
  \hline
  Time slot duration & $T_d$\\
  \hline
  Maximum UAV speed & $V_{\rm max}$\\
  \hline
  Path loss exponent for LoS & $\alpha_L$\\
  \hline
  Path loss exponent for NLoS & $\alpha_N$\\
  \hline
  Rician factor & $\kappa$\\
  \hline
  Channel coefficient & $h$\\
  \hline
  Channel rate & $C$\\
  \hline
  $m$-th EHBS transmit power & $P_m$\\
  \hline
  Distance & $d$\\
  \hline
  Bandwidth & $B$\\
  \hline
  Number of tasks & $K$\\
  \hline
  Generation quality for $k$-th task & $R_k$\\
  \hline
  Latency for $k$-th task & $T_k$\\
  \hline
  Generation quality threshold for $k$-th task & $R_k^{\min}$\\
  \hline
  Latency threshold for $k$-th task & $T_k^{\max}$\\
  \hline
  EHBS selection indicator & $S_m$\\
  \hline
  User association indicator & $V_n$\\
  \hline
  Linear energy harvesting efficiency  & $\alpha$\\
  \hline
  Non-Linear energy harvesting factor  & $\xi$\\
  \hline
  Non-Linear energy harvesting factor  & $\epsilon$\\
  \hline
  Maximum harvested power at receiver & $W$\\
  \hline
  Conventional received RF power & $P_{{\rm RF},m}(t)$\\
  \hline
  Energy harvesting time factor  & $\beta$\\
  \hline
  Generation quality weight & $\rho_q$\\
  \hline
  Latency weight  & $\rho_d$\\
  \hline
  SEM for $k$-th task  & $\mathcal{E}_k$\\
  \hline
  UAV maximum altitude  & $z_{\max}$\\
  \hline
  UAV minimum altitude  & $z_{\min}$\\
  \hline
  Discount factor  & $\gamma$\\
  \hline
  State at time slot $t$  & $f(t)$\\
  \hline
  Action at time slot $t$  & $a(t)$\\
  \hline
  Reward at time slot $t$  & $r(t)$\\
  \hline
  Reward constraint factor & $\delta$\\
  \hline
 \end{tabular}
 \label{tab:pares}
 \end{table}

\section{System Model}\label{sec:sm}
\begin{figure}[t!]
  \centering
  \centerline{\includegraphics[width=0.47\textwidth]{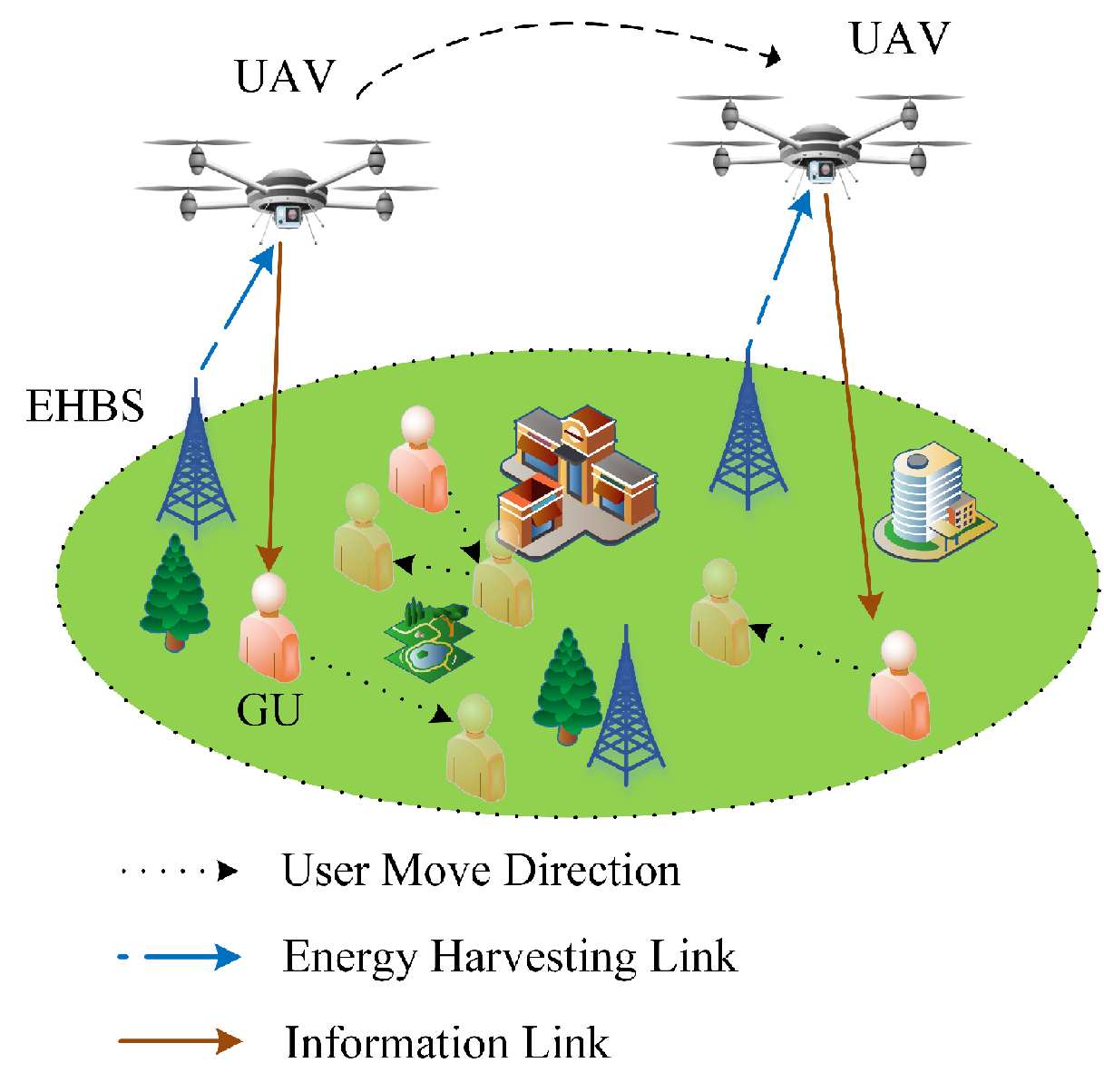}}
 \caption{System model of UAV-enabled WPT networks.} \label{fig:SM}
\end{figure}
We consider a UAV network assisted by $M$ ground energy-harvesting base stations (EHBSs) and serving $N$ ground users (GUs), where $EH_m$ denotes the $m$-th EHBS, $m\in\mathcal{M} = \{1,\ldots,M\}$, and $G_n$ denotes the $n$-th GU, $n\in\mathcal{N} = \{1,\ldots,N\}$. The total service duration is slotted, indexed by $t\in\mathcal{T} = \{1,\ldots,T\}$, and each slot has duration $T_d$. The UAV $U$ flies at the 3D coordinate $\mathbf{q}(t)=\big[x(t),\,y(t),\,z(t)\big]$, while the GUs follow the random waypoint model. The UAV kinematics satisfy
\begin{equation}
\|\mathbf{q}(t{+}1)-\mathbf{q}(t)\|\le v_{\max}T_d,\quad z_{\min}\le z(t)\le z_{\max}, \label{eq:mobility}
\end{equation}
where $v_{\rm \max}$ is the maximum speed. At each slot, the UAV selects one EHBS for wireless power transfer (WPT) and serves one GU for downlink information transmission. It is important to mention that, unlike many schemes that choose the strongest EHBS, we do not simply select the strongest EHBS in this work. This is because the UAV trajectory affects the downlink transmission performance to the ground user. When trajectory and EHBS selection are jointly optimized, the strongest is not a fixed choice.

Ground users randomly generate transmission tasks, and the UAV is responsible for transmitting data to the corresponding users. We adopt the Rician fading model for UAV-GU channels \cite{8698468} as
\begin{equation}\label{eq:rician_u2g}
h_{u,n}(t)= \sqrt{\frac{\kappa}{\kappa+1}}\,\bar{H}_{u,n}(t)\;+\;\sqrt{\frac{1}{\kappa+1}}\,\hat{H}_{u,n}(t),
\end{equation}
where the line of sight (LoS) and non-line-of-sight (NLoS) components are
\begin{equation}
\bar{H}_{u,n}(t)=\bar{\mathbb{g}}_{u,n}\, d_{u,n}(t)^{-\frac{\alpha_L}{2}},
\end{equation}
and
\begin{equation}
\hat{H}_{u,n}(t)=\hat{\mathbb{g}}_{u,n}\, d_{u,n}(t)^{-\frac{\alpha_N}{2}},
\end{equation}
respectively. Here, $\kappa$ is the Rician factor; $\alpha_L$ and $\alpha_N$ denote the path-loss exponents for LoS and NLoS components, respectively. $\bar{\mathbb{g}}_{u,n}$ is a deterministic unit-modulus coefficient for LoS component, and $\hat{\mathbb{g}}_{u,n}$ follows zero-mean unit-variance Gaussian distribution. The WPT links for UAV-EHBS are modeled the same as UAV-GU.

The communication process is divided into two stages. In the first stage, the UAV selects one EHBS to harvest energy, let a binary indicator $S_m \in \{0, 1\}$ denotes the selection $S_m = 1$ means $m$-th EHBS is selected for energy harvesting, otherwise $S_m = 0$. In the second stage, the UAV transmits data to a ground user, where the UAV’s transmit power comes from the energy harvested in the first stage. We assume a binary indicator $V_n \in \{0, 1\}$ for user association, where $V_n = 1$ means $n$-th GU is selected for transmission in this stage, otherwise $V_n = 0$. The time durations of the first and second stages are $\beta T_d$ and $(1-\beta) T_d$, respectively, where $\beta \in (0, 1)$ denotes the energy harvesting time factor. Thus, each slot $t$ follows a harvest-then-transmit structure. Let $EH_m(t)$ denote the energy harvesting EHBS at time slot $t$, based on the linear energy harvesting model, we can obtain the UAV’s transmit power in the second stage is \cite{7320989}
\begin{equation}\label{eq:ptx}
P_u(t)= \frac{\alpha \beta(t)}{1-\beta(t)}\, P_{m}(t) \big|h_{u,m}(t)\big|^2,
\end{equation}
where $\alpha$ denotes the WPT efficiency from EHBS, $P_{m}(t)$ is the transmit power at EHBS $EH_m$. However, the power conversion efficiency is a key performance indicator in practical RF energy harvesting circuits. Many experimental studies have verified that the relationship between the input and output power is strongly nonlinear. Thus, different from the conventional energy harvesting linear model, we introduce the non-linear wireless power transfer model as \cite{7843670}. Thus, the harvested power at UAV can be expressed as
\begin{equation}\label{eq:nonlinear_eh}
P_{u}(t)=\frac{\beta(t)}{1-\beta(t)}\frac{\Psi_{m}(t)-W \varsigma}{1-\varsigma},
\end{equation}
where
\begin{equation}
\Psi_{m}(t)=\frac{W}{1+\exp\left(-\xi\left(P_{{\rm RF},m}(t)-\epsilon\right)\right)},
\end{equation}
and
\begin{equation}
\varsigma=\frac{1}{1+\exp(\xi \epsilon)}, P_{{\rm RF},m}(t)=P_{m}(t){|h_{u,m}(t)|}^2,
\end{equation}
where $W$ denotes the maximum harvested power for the UAV, $\xi$ and $\epsilon$ are constants of energy harvesting circuit physical hardware, $P_{{\rm RF},m}(t)$ is the conventional logistic function for the received RF power.

The UAV selects one GU to transmit data at one time slot. Let the scheduled user at slot $t$ be denoted by $G_{n}$. During the second transmission stage, the received signal at $G_{n}$ is
\begin{equation}\label{eq:rx_signal}
y_{n}(t) \;=\; \sqrt{P_u(t)}\, h_{u, n}(t)\, x(t) + n_n(t),
\end{equation}
where $x(t)$ is a transmit signal from the UAV, and $n_n(t)$ denotes additive white Gaussian noise (AWGN) at $G_{n}$. Thus, the instantaneous SNR at time slot $t$ between UAV and $G_{n}$ is
\begin{equation}\label{eq:snr_def}
SNR_{n}(t)\;=\;\frac{P_u(t)\,\big|h_{u, n}(t)\big|^2}{\sigma^2}.
\end{equation}
Therefore, the channel capacity for the link between UAV $U$ and $G_{n}$ is given by
\begin{equation}\small\label{8a}
   C_{u,n} = B(1-\beta){\rm{log_{2}}}\left( 1+ SNR_{n}(t) \right),
\end{equation}
where $B$ denotes the bandwidth for UAV-GU transmissions.

\section{Hybrid Bit \& Semantic Communication Framework and Problem Formulation} \label{sec:semantic}
To avoid collisions with previously used symbols, we adopt the following notation throughout this section. The raw image is denoted by $\mathbf{r}$. A text embedding extracted from a caption is $\omega$. The quantized visual semantic representation is $\mathbf{c}$. Diffusion latents are $\mathbf{y}_\tau$ at step $\tau\!\in\!\{1,\dots,\Theta\}$, the injected noise is $\boldsymbol{\nu}$, and the noise scheduler uses a sequence $\{\eta_\tau\}_{\tau=1}^{\Theta}$ with $\varphi_\tau\!=\!1-\eta_\tau$ and $\bar{\varphi}_\tau\!=\!\prod_{i=1}^{\tau}\varphi_i$. The denoiser network (UNet) parameters are $\psi$. For task-level evaluation, we use $\{\mathcal{T}_k\}_{k=1}^{K}$ with score $\mathcal{E}_k$, normalized reconstruction quality $R_k\!\in\![0,1]$, normalized end-to-end latency $T_k\!\in\![0,1]$, minimum required quality $R_k^{\min}$, latency threshold $T_k^{\max}$, and weights $\rho_q,\rho_d\!\in\!(0,1)$ with $\rho_q+\rho_d=1$.

\subsection{Hybrid Bit \& Semantic Communication Framework}
\begin{figure*}[t!]
    \centering
    \includegraphics[width=1\textwidth]{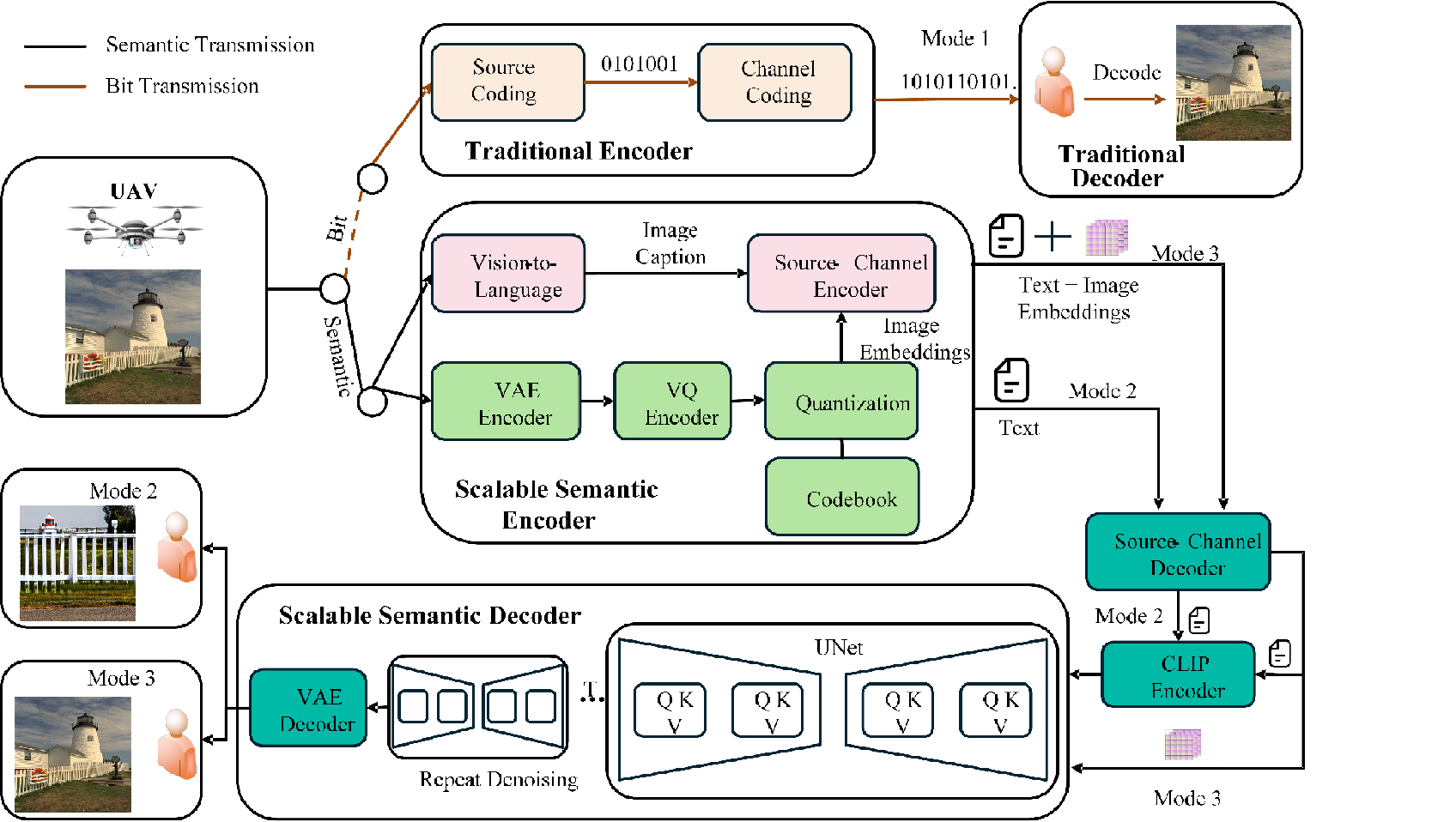}
    \caption{The Hybrid Bit \& Semantic Communication Framework.}
    \label{semanticframework}
\end{figure*}
To serve different link budgets in UAV communications, the proposed framework exposes three switchable operating modes as shown in Fig. \ref{semanticframework}:
\begin{itemize}
    \item \textbf{Mode-$\boldsymbol{1}$ (Bit transmission).} A classic source-channel transmission. It provides high image quality at the receiver but requires large bit rates and limited scalability, especially under low SNR. JPEG is used as the traditional compression method in this mode.
    \item \textbf{Mode-$\boldsymbol{2}$ (Text features transmission).} Only a textual description of the image is transmitted; the receiver reconstructs the image from the text. This offers the highest compression but typically lower reconstruction fidelity and higher compute demand at the decoder.
    \item \textbf{Mode-$\boldsymbol{3}$ (Text + visual features transmission).} Both a caption embedding and a compact, quantized visual representation are sent. By varying the spatial resolution and the codebook size, the bitrate--quality trade-off can be adjusted continuously. In this work, we designed three sub-modes for Mode 3, each corresponding to different levels of compression capability and semantic generation quality.
\end{itemize}

The encoder comprises a vision-to-language front-end, a variational autoencoder (VAE)~\cite{vae}, a vector-quantization encoder (VQE), a shared codebook, and a source--channel encoder \cite{11207608}:
\begin{enumerate}
    \item \textbf{Text branch.} BLIP-2~\cite{blip2} produces a caption, which is mapped by a CLIP text encoder~\cite{clip} to the embedding $\omega$.
    \item \textbf{Vision branch.} The raw image $\mathbf{r}$ is mapped by the VAE into a latent feature map. This latent is then processed by the VQE:
    \begin{itemize}
        \item stacked convolutional, residual, and attention blocks distill high-level semantics;
        \item the spatial size is projected to $l\times u$ according to the target bitrate (stride-$2$ convolutions when downsampling is required; otherwise $1\times1$ convolutions);
        \item a gated attention mechanism modulates channels by learned gates;
        \item after $N_b$ stages, vector quantization is applied against a codebook, so we can obtain the quantized representation $\mathbf{c}$.
    \end{itemize}
\end{enumerate}
If the codebook cardinality is $L$, the visual payload scales as approximately $l u\log_2 L$ bits. Thus, Mode-3 can finely tune bitrate and fidelity via $(l,u,L)$. UAVs in the aerial layer can further relay the semantic payloads to users when required by resource constraints.

The receiver reconstructs images by running a conditional diffusion process guided by $\omega$ and $\mathbf{c}$.
Given the scheduler $\{\eta_\tau\}$ with $\varphi_\tau=1-\eta_\tau$ and $\bar{\varphi}_\tau=\prod_{i=1}^{\tau}\varphi_i$, the forward noising procedure is
\begin{equation}
\mathbf{y}_\tau
= \sqrt{\bar{\varphi}_\tau}\,\mathbf{y}_0
+ \sqrt{1-\bar{\varphi}_\tau}\,\boldsymbol{\nu},
\qquad \boldsymbol{\nu}\sim\mathcal{N}(\mathbf{0},\mathbf{Id}),
\end{equation}
where $\mathbf{y}_0$ denotes a clean latent and $\mathbf{Id}$ is the identity matrix.
During reverse denoising, a UNet~\cite{sdm} predicts step-wise noise and updates
\begin{equation}
\label{eq:rev}
\mathbf{y}_{\tau-1}
= \frac{1}{\sqrt{\varphi_\tau}}\!\left(
\mathbf{y}_\tau
- \frac{1-\varphi_\tau}{\sqrt{1-\bar{\varphi}_\tau}}\,
\hat{\nu}_{\psi}\!\left(\mathbf{y}_\tau,\tau,\mathbf{c},\omega\right)
\right).
\end{equation}
Implementation-wise, the visual embedding $\mathbf{c}$ is concatenated with the current noisy latent along the batch dimension to form a hybrid tensor, and cross-attention layers inject global textual constraints from $\omega$ while preserving local priors from $\mathbf{c}$.
Iterating~\eqref{eq:rev} down to $\tau=0$ yields $\mathbf{y}_0$, which is decoded by the VAE to produce the final image.

The VAE, BLIP-2, and CLIP are adopted from strong public models and kept frozen to provide stable representations and reduce training cost. Following the VQ-VAE paradigm~\cite{vq-vae}, each local feature vector $\mathbf{p}$ is mapped to its nearest codeword $\mathbf{c}_q$ in the codebook. Using $\operatorname{stop}(\cdot)$ to block gradients, the VQ loss is
\begin{equation}
\mathcal{L}_{\mathrm{vq}}
=\mathbb{E}_{\mathbf{p}}\!\left[
\big\|\operatorname{stop}(\mathbf{p})-\mathbf{c}_q\big\|_2^2
+ \big\|\operatorname{stop}(\mathbf{c}_q)-\mathbf{p}\big\|_2^2
\right].
\end{equation}
Since transmitter and receiver share the same codebook, transmitting only codeword indices suffices for reconstructing $\mathbf{c}$ at the decoder, significantly reducing bandwidth.

With $\boldsymbol{\nu}\!\sim\!\mathcal{N}(\mathbf{0},\mathbf{Id})$, the UNet is trained by mean-squared error:
\begin{equation}
\mathcal{L}_{\mathrm{denoiser}}
=\mathbb{E}_{P_{\mathbf{r}}}\,
\mathbb{E}_{P_{\mathbf{c},\mathbf{y}_\tau|\mathbf{r}}}\,
\mathbb{E}_{\boldsymbol{\nu}}\,
\big\|
\boldsymbol{\nu}
- \hat{\nu}_{\psi}(\mathbf{y}_\tau,\tau,\mathbf{c},\omega)
\big\|_2^2.
\end{equation}

\begin{figure*}[t!]
  \centering
  \centerline{\includegraphics[width=0.96\textwidth]{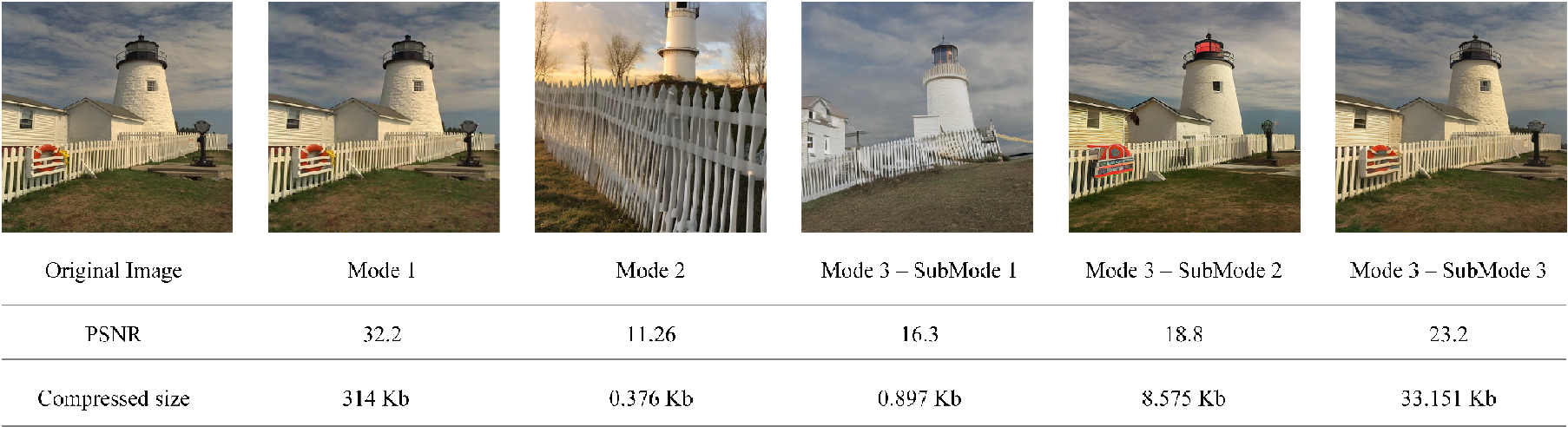}}
 \caption{The compressed size and semantic generation quality for different transmission modes.} \label{fig:semanticModes}
\end{figure*}

We employ classifier-free guidance (CFG) \cite{ho2021classifierfree} at inference to modulate the effect of text conditioning with a guidance factor $\omega_g$:
\begin{equation}
\begin{aligned}
\tilde{\nu}_{\psi}
= ~&\hat{\nu}_{\psi}\!\big(\mathbf{y}_\tau,(\mathbf{c},\varnothing),\tau\big)
+ \omega_g\! \bigg[
\hat{\nu}_{\psi}\!\big(\mathbf{y}_\tau,(\mathbf{c},\omega),\tau\big) \\
&- \hat{\nu}_{\psi}\!\big(\mathbf{y}_\tau,(\mathbf{c},\varnothing),\tau\big)
\bigg].
\end{aligned}
\end{equation}
To make CFG feasible, text conditions are randomly dropped with probability $p_{\varnothing}\!\approx\!0.1$ during training and replaced by learnable null embeddings. The compressed size and semantic generation quality for each semantic transmission mode can be found in Fig. \ref{fig:semanticModes}.

\subsection{Semantic Communication Efficiency Metric}
We introduce the Semantic Efficiency Metric (SEM) to jointly evaluate latency and reconstruction quality under tight link budgets in UAV communications. For a set of $K$ image transmission tasks $\{\mathcal{K}\}$, we define
\begin{equation}\label{eq:sce}
\mathcal{E}_k
= \rho_q\big(R_k - R_k^{\min}\big)
+ \rho_d\big(T_k^{\max} - T_k\big),
\end{equation}
where $R_k$ is the normalized reconstruction metric (e.g., peak signal to noise ratio (PSNR)) at the receiver, and $T_k$ is the normalized transmission latency for $k$-th task; $R_k^{\min}$ and $T_k^{\max}$ are task-specific thresholds for minimum acceptable quality and maximum allowable latency, respectively. Eq. \eqref{eq:sce} provides a single scalar objective for mode selection and rate allocation: when thresholds are tight or links are weak (small $T_k^{\max}$ or large $T_k$), the system prefers Mode-2 or a lower-rate configuration of Mode-3. When quality is paramount (large $R_k^{\min}$), the system raises the semantic strength in Mode-3 or falls back to Mode-2 and Mode-1 for maximal generation quality.

In summary, the proposed design couples a three-mode, rate-scalable hybrid pipeline (Mode 1, Mode 2 and Mode 3) with VAE--VQE semantic compression and conditional diffusion reconstruction. The SEM metric in~\eqref{eq:sce} offers a principled way to navigate latency-quality trade-offs in UAV communications. Based on the resource budget in UAV networks, different semantic transmission modes show different suitability. Mode-$\boldsymbol{1}$ generally requires the highest bit rate and is more appropriate when the harvested energy is sufficient and the channel condition is favorable. Mode-$\boldsymbol{2}$ has the smallest payload requirement and is thus more suitable when the communication resource is limited, such as under weak channels, low harvested energy, or tight latency constraints. Mode-$\boldsymbol{3}$ provides a flexible intermediate solution, since its visual semantic payload can be adjusted through the spatial resolution and codebook size. The hybrid transmission framework allows the UAV to adapt the transmission mode according to the currently available resource budget rather than relying on a single fixed transmission strategy, this is the reason we introduce this framework to the UAV communications.

\subsection{Problem Formulation}
In the proposed UAV-enabled WPT networks, we aim to maximize the average semantic transmission efficiency SEM. We can optimize the UAV 3D trajectory, EHBS selection, user association, hybrid transmission mode selection, and energy harvesting time factor in this scenario. Therefore, the problem formulation can be expressed as
\begin{align}
    \bold{\rm (P1)}: &\max_{q(t), {\mathcal S}(t), {\mathcal V}(t), \mathcal{O}(t), \beta(t)} \frac{1}{K}\sum_{k=1}^{K} \mathcal{E}_k,\label{SecrecyFunc}\\
    {\rm s.t.}&~ \sum_{m=1}^{M} S_{m} = 1 \tag{\ref{SecrecyFunc}{a}}, \label{SecrecyFuncSuba}\\
    &\sum_{n=1}^{N} V_{n} = 1 \tag{\ref{SecrecyFunc}{b}}, \label{SecrecyFuncSubb}\\
    &\|\mathbf{q}(t{+}1)-\mathbf{q}(t)\|\le v_{\rm \max}T_d \tag{\ref{SecrecyFunc}{c}}, \label{SecrecyFuncSubc}\\
    &\quad z_{\min}\le z(t)\le z_{\max} \tag{\ref{SecrecyFunc}{d}}, \label{SecrecyFuncSubd}\\
    &\beta \in (0, 1) \tag{\ref{SecrecyFunc}{e}}, \label{SecrecyFuncSube}\\
    &T_k \leq T_k^{\max}, \forall k \in {\mathcal K} \tag{\ref{SecrecyFunc}{f}}, \label{SecrecyFuncSubf}\\
    &R_k \geq R_k^{\min}, \forall k \in {\mathcal K} \tag{\ref{SecrecyFunc}{g}}, \label{SecrecyFuncSubg}
\end{align}
where ${\mathcal S}(t) = \{S_1, S_2, ..., S_M\}$ denotes the EHBS selection, ${\mathcal V}(t)= \{V_1, V_2, ..., V_N\}$ denotes the GU selection, $\mathcal{O}(t) = \{O_1, O_2, O_{3,1}, O_{3,2}, O_{3,3}\}$ denotes the hybrid transmission mode selection, $O_1$ is Mode 1, $O_2$ is Mode 2, $O_{3,1}$, $O_{3,2}$ and $O_{3,3}$ are the sub-modes of Mode 3. \eqref{SecrecyFuncSuba} indicates that only one EHBS can be selected at time slot $t$, \eqref{SecrecyFuncSubb} shows that one GU can be selected to receive image from the UAV. \eqref{SecrecyFuncSubc} and \eqref{SecrecyFuncSubd} present the mobility limitations of the UAV. \eqref{SecrecyFuncSube} shows the constraint of energy harvesting time factor. \eqref{SecrecyFuncSubf} and \eqref{SecrecyFuncSubg} indicate that the transmission tasks need to satisfy the threshold requirements. Considering that the proposed UAV communication network involves time-varying Rician fading channels, UAV trajectory variations, and uncertainty in transmission tasks, the formulated problem belongs to a long-term optimization problem in a dynamic network. Therefore, to obtain fast decisions in such a highly dynamic environment and to maximize the long-term performance, we adopt a DRL algorithm to maximize the semantic efficiency metric.

\section{DRL-Based Resource Allocation} \label{sec:DSAC}
To solve this long-term planning problem in the dynamic UAV network by using DRL, we first model the problem as a Markov decision process (MDP). Therefore, we define the state as
\begin{equation}\small\label{eq:state}
\begin{aligned}
f(t) =&~\{t, {h_{u, n}(t)}_{n \in \mathcal N}, {h_{u, m}(t)}_{m \in \mathcal M}, q(t-1), \\
&{R_k(t)}_{k \in \mathcal K}, {T_k(t)}_{k \in \mathcal K}\}.
\end{aligned}
\end{equation}
Then, we assume the action at time slot $t$ is
\begin{equation}\small\label{eq:action}
\begin{aligned}
a(t) = \{ a_h(t), a_v(t), v(t), {\mathcal S}(t), {\mathcal O}(t), \beta(t) \},
\end{aligned}
\end{equation}
where $a_h(t)$ and $a_v(t)$ represent the UAV's horizontal flight angle and vertical flight angle, respectively. $v(t)$ is the UAV speed. Notice that the angles and speed are discretized into ten uniform intervals. Moreover, the reward function plays a key role in training the DRL agent, serving as the feedback signal from the environment to enhance the policy performance of the agent in the learning process. We define the reward function as
\begin{equation}\small\label{eq:reward}
\begin{aligned}
r(t) = \sum_{k=1}^{K} (\mathcal{E}_k - \delta),
\end{aligned}
\end{equation}
where $\delta = 1$ when any constraints are not satisfied, otherwise $\delta = 0$.

To learn a resource-allocation policy that maximizes the cumulative semantic communication efficiency while explicitly modeling uncertainty in long-horizon returns, we adopt a distributional soft actor-critic (DSAC) formulation \cite{9448360}. Unlike conventional SAC \cite{11270936}, which backs up the mean of entropy-regularized returns, DSAC fits and propagates a parametric distribution of soft returns via a distributional Bellman operator.

Let $\mathcal{P}$ denote the current stochastic policy, $\boldsymbol{x}$ a state, and $\boldsymbol{u}$ an action. We define the one-step soft target as
\begin{equation}\label{eq:dsacA1}
\mathsf{T}^{\mathcal{P}}(\boldsymbol{x},\boldsymbol{u})
=
r + \gamma\,\mathcal{G}(\boldsymbol{x}^{+}),
\end{equation}
where $r$ is the immediate reward, $\gamma\in(0,1)$ the discount factor, and $\mathcal{G}(\boldsymbol{x}^{+})$ the entropy-regularized return-to-go at the successor state $\boldsymbol{x}^{+}$.

The soft state-action value equals the expectation of \eqref{eq:dsacA1}:
\begin{equation}\label{eq:dsacA2}
\mathcal{Q}^{\mathcal{P}}(\boldsymbol{x},\boldsymbol{u})
=
\mathbb{E}\!\left[\mathsf{T}^{\mathcal{P}}(\boldsymbol{x},\boldsymbol{u})\right].
\end{equation}
Instead of collapsing to the mean in \eqref{eq:dsacA2}, DSAC propagates a distribution using the distributional soft Bellman operator
\begin{equation}\label{eq:dsacA3}
\mathcal{T}_{\mathrm{soft}}\!\left[\mathsf{Z}^{\mathcal{P}}(\boldsymbol{x},\boldsymbol{u})\right]
=
r + \gamma\Big(
\mathsf{Z}^{\mathcal{P}}(\boldsymbol{x}^{+},\boldsymbol{u}^{+})
-
\alpha\,\log \mathcal{P}(\boldsymbol{u}^{+}\mid\boldsymbol{x}^{+})
\Big),
\end{equation}
where $\mathsf{Z}^{\mathcal{P}}(\boldsymbol{x},\boldsymbol{u})$ is a random variable for the soft return and $\boldsymbol{u}^{+}\sim\mathcal{P}(\cdot\mid\boldsymbol{x}^{+})$. The scalar $\alpha>0$ is the entropy temperature.

Let $\widehat{\mathcal{Z}}_{\boldsymbol{\theta}}(\cdot\mid\boldsymbol{x},\boldsymbol{u})$ be a parametric density with parameters $\boldsymbol{\theta}$, and let $\bar{\boldsymbol{\theta}}$ denote a slow-moving target copy. At each update we project the pushed-forward target distribution $\mathcal{T}_{\mathrm{soft}}\widehat{\mathcal{Z}}_{\bar{\boldsymbol{\theta}}}$ onto $\widehat{\mathcal{Z}}_{\boldsymbol{\theta}}$ by minimizing a divergence (here, KL):
\begin{equation}\label{eq:dsacA4}
\widehat{\mathcal{Z}}_{\boldsymbol{\theta}}^{\mathrm{new}}
=
\arg\min_{\widehat{\mathcal{Z}}_{\boldsymbol{\theta}}}
\;
\mathbb{E}\!\left[
\mathbb{D}_{\mathrm{KL}}\!\left(
\mathcal{T}_{\mathrm{soft}}\widehat{\mathcal{Z}}_{\bar{\boldsymbol{\theta}}}(\cdot\mid\boldsymbol{x},\boldsymbol{u})
\;\Big\|\;
\widehat{\mathcal{Z}}_{\boldsymbol{\theta}}(\cdot\mid\boldsymbol{x},\boldsymbol{u})
\right)
\right].
\end{equation}

A practical negative log-likelihood objective that realizes \eqref{eq:dsacA4} is
\begin{equation}\label{eq:dsacA5}
\begin{aligned}
\mathcal{L}_{\mathcal{Z}}(\boldsymbol{\theta})
=~&
-
\mathbb{E}_{(\boldsymbol{x},\boldsymbol{x}^{+},\boldsymbol{u},r)\sim\mathcal{D}}
\;
\mathbb{E}_{\boldsymbol{u}^{+}\sim\mathcal{P}_{\boldsymbol{\phi}}(\cdot\mid\boldsymbol{x}^{+})}
\;
\mathbb{E}_{z^{+}\sim \widehat{\mathcal{Z}}_{\bar{\boldsymbol{\theta}}}(\cdot\mid\boldsymbol{x}^{+},\boldsymbol{u}^{+})}\\
&\left[
\log q_{\boldsymbol{\theta}}\!\big(\Phi(z^{+};\boldsymbol{x},\boldsymbol{u})\big)
\right],
\end{aligned}
\end{equation}
where $\mathcal{D}$ is a replay buffer, $q_{\boldsymbol{\theta}}$ is the density of $\widehat{\mathcal{Z}}_{\boldsymbol{\theta}}$, $\boldsymbol{\phi}$ are the policy parameters used to sample $\boldsymbol{u}^{+}$, and $\Phi(\cdot)$ transports a target draw through \eqref{eq:dsacA3}. Taking gradients yields
\begin{equation}\label{eq:dsacA6}
\nabla_{\boldsymbol{\theta}}\mathcal{L}_{\mathcal{Z}}(\boldsymbol{\theta})
=
-
\mathbb{E}\!\left[
\nabla_{\boldsymbol{\theta}}\log q_{\boldsymbol{\theta}}\!\big(\Phi(z^{+};\boldsymbol{x},\boldsymbol{u})\big)
\right].
\end{equation}

Assume $\widehat{\mathcal{Z}}_{\boldsymbol{\theta}}(\cdot\mid\boldsymbol{x},\boldsymbol{u})
=
\mathcal{N}\!\big(\mu_{\boldsymbol{\theta}}(\boldsymbol{x},\boldsymbol{u}),\,\sigma_{\boldsymbol{\theta}}(\boldsymbol{x},\boldsymbol{u})^{2}\big)$.
For $\tilde{z}=\Phi(z^{+};\boldsymbol{x},\boldsymbol{u})$, the score-function derivatives of the log-density are
\begin{equation}\label{eq:dsacA7}
\frac{\partial \log q_{\boldsymbol{\theta}}(\tilde{z})}{\partial \mu_{\boldsymbol{\theta}}}
=
\frac{\tilde{z}-\mu_{\boldsymbol{\theta}}}{\sigma_{\boldsymbol{\theta}}^{2}},
\qquad
\frac{\partial \log q_{\boldsymbol{\theta}}(\tilde{z})}{\partial \sigma_{\boldsymbol{\theta}}}
=
\frac{(\tilde{z}-\mu_{\boldsymbol{\theta}})^{2}}{\sigma_{\boldsymbol{\theta}}^{3}}
-
\frac{1}{\sigma_{\boldsymbol{\theta}}}.
\end{equation}
To avoid ill-conditioned gradients when $\sigma_{\boldsymbol{\theta}}$ is too small or too large, we enforce a minimum scale
\begin{equation}\label{eq:dsacA8}
\sigma_{\boldsymbol{\theta}}(\boldsymbol{x},\boldsymbol{u})
\leftarrow
\max\!\big(\sigma_{\boldsymbol{\theta}}(\boldsymbol{x},\boldsymbol{u}),\,\underline{\sigma}\big),
\end{equation}
with hyperparameter $\underline{\sigma}>0$. We further apply target clipping to mitigate overestimation:
\begin{equation}\label{eq:dsacA9}
\tilde{z}
\leftarrow
\operatorname{clip}\!\Big(
\tilde{z},\,
\mu_{\boldsymbol{\theta}}(\boldsymbol{x},\boldsymbol{u})-\varepsilon,\,
\mu_{\boldsymbol{\theta}}(\boldsymbol{x},\boldsymbol{u})+\varepsilon
\Big),
\end{equation}
where $\varepsilon>0$ sets the window width.

Given the learned return model, the actor maximizes
\begin{equation}\label{eq:dsacA10}
\mathcal{J}_{\pi}(\boldsymbol{\phi})
=
\mathbb{E}_{\boldsymbol{x}\sim\mathcal{D},\,\boldsymbol{u}\sim\mathcal{P}_{\boldsymbol{\phi}}(\cdot\mid\boldsymbol{x})}
\left[
\underbrace{\mathcal{Q}_{\boldsymbol{\theta}}(\boldsymbol{x},\boldsymbol{u})}_{\text{mean of }\widehat{\mathcal{Z}}_{\boldsymbol{\theta}}}
-
\alpha\,\log \mathcal{P}_{\boldsymbol{\phi}}(\boldsymbol{u}\mid\boldsymbol{x})
\right].
\end{equation}
Using the re-parameterization $\boldsymbol{u}=g_{\boldsymbol{\phi}}(\boldsymbol{x},\varepsilon)$ with fixed noise $\varepsilon$, the policy gradient is
\begin{equation}\label{eq:dsacA11}
\begin{aligned}
\nabla_{\boldsymbol{\phi}}\mathcal{J}_{\pi}(\boldsymbol{\phi})
=~&
\mathbb{E}_{\boldsymbol{x},\varepsilon}
\bigg[
\nabla_{\boldsymbol{\phi}} g_{\boldsymbol{\phi}}(\boldsymbol{x},\varepsilon)\,
\Big(
\nabla_{\boldsymbol{u}}\mathcal{Q}_{\boldsymbol{\theta}}(\boldsymbol{x},\boldsymbol{u})\\
&
-
\alpha\,\nabla_{\boldsymbol{u}}\log \mathcal{P}_{\boldsymbol{\phi}}(\boldsymbol{u}\mid\boldsymbol{x})
\Big)
\bigg].
\end{aligned}
\end{equation}

Equations \eqref{eq:dsacA1}-\eqref{eq:dsacA11} summarize the proposed DSAC variant. We fit a Gaussian return head via KL projection with a minimum-scale constraint and clipped targets, then improve the actor under the maximum-entropy principle using a re-parameterized sampler. This aligns the learned distribution of soft returns with entropy-regularized control while improving numerical stability during training.

\subsection{Decision-Assisted DRL}

In our wireless-powered UAV semantic communication network, agents must respect both wireless communication reachability and power transfer constraints. In the previous section, the reward shaping adopted for DSAC assigns a payoff to every action without first checking whether the action is feasible under the current state (e.g., link budget, UAV location, GU locations, task status, harvested energy). This assumption is problematic in the proposed network. For example, if the UAV $U$ has already finished a specific transmission task, then continuing to allocate transmission link for this specific task is an unavailable action. Likewise, when an EHBS has a very weak link between itself and the UAV, selecting this EHBS at the first energy harvesting stage is also unavailable. If such actions are not filtered, the DRL agent must spend many updates to infer the mutual influences among state–action pairs, which reduces sample efficiency and may even prevent convergence.

A common remedy in DRL is to give negative rewards to ineffective or goal-irrelevant actions, thereby biasing exploration away from them and accelerating learning. However, this purely punitive strategy can be overly harsh during early training. With randomly initialized parameters, the function approximator may output small or near-zero values even for the optimal action over several iterations; strong penalties can then steer the policy toward suboptimal regions and increase the risk of convergence to poor local optima \cite{10440193}. Another straightforward idea is to hard-ban infeasible actions so they cannot be selected. Yet, in most parameterizations the same policy network produces preferences for all actions. Even if some choices are filtered at execution time, their logits (for discrete) or continuous outputs still backpropagate, reshaping the shared representation and allowing infeasible actions to distort what the network learns.

\begin{figure}[t!]
  \centering
  \centerline{\includegraphics[width=0.43\textwidth]{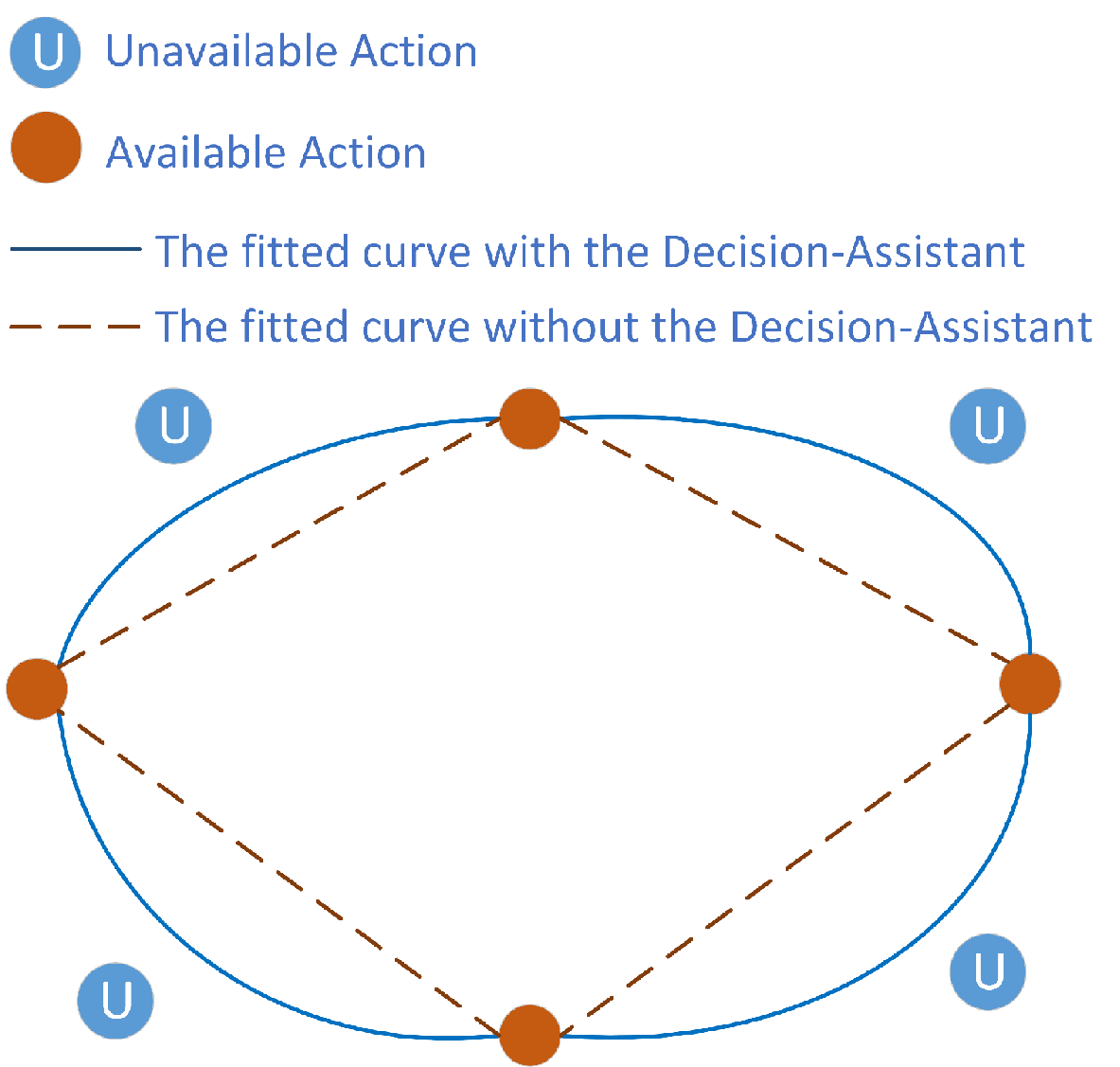}}
 \caption{Unavailable action's impact on DRL training.} \label{fig:UA}
\end{figure}

This effect is qualitatively illustrated in Fig. \ref{fig:UA}. If unavailable actions are simply ignored, the fitted mapping behaves as if it “connects’’ only feasible choices (dashed curves). When unavailable actions are present and penalized, the regressor is pushed away from those points, shifting the fitted surface (solid curves). In both cases, unavailable actions can perturb the function approximator and indirectly affect the resource-allocation decisions of other UAVs.

To mitigate this issue, we adopt a decision-assisted DRL scheme that injects feasibility knowledge directly into training \cite{9321455}. Rather than rewarding or sampling unavailable actions, we assign fixed minimal preferences to state–action pairs that violate the feasibility rules, so that they contribute as little as possible to the gradient signal. Concretely, for each mini-batch we label sampled state–action pairs with a feasibility flag determined by simple tests. Feasible actions are optimized by the standard SAC objective, while infeasible actions are driven toward a fixed baseline output so that their influence on the shared representation is suppressed.

\begin{figure}[t!]
  \centering
  \centerline{\includegraphics[width=0.45\textwidth]{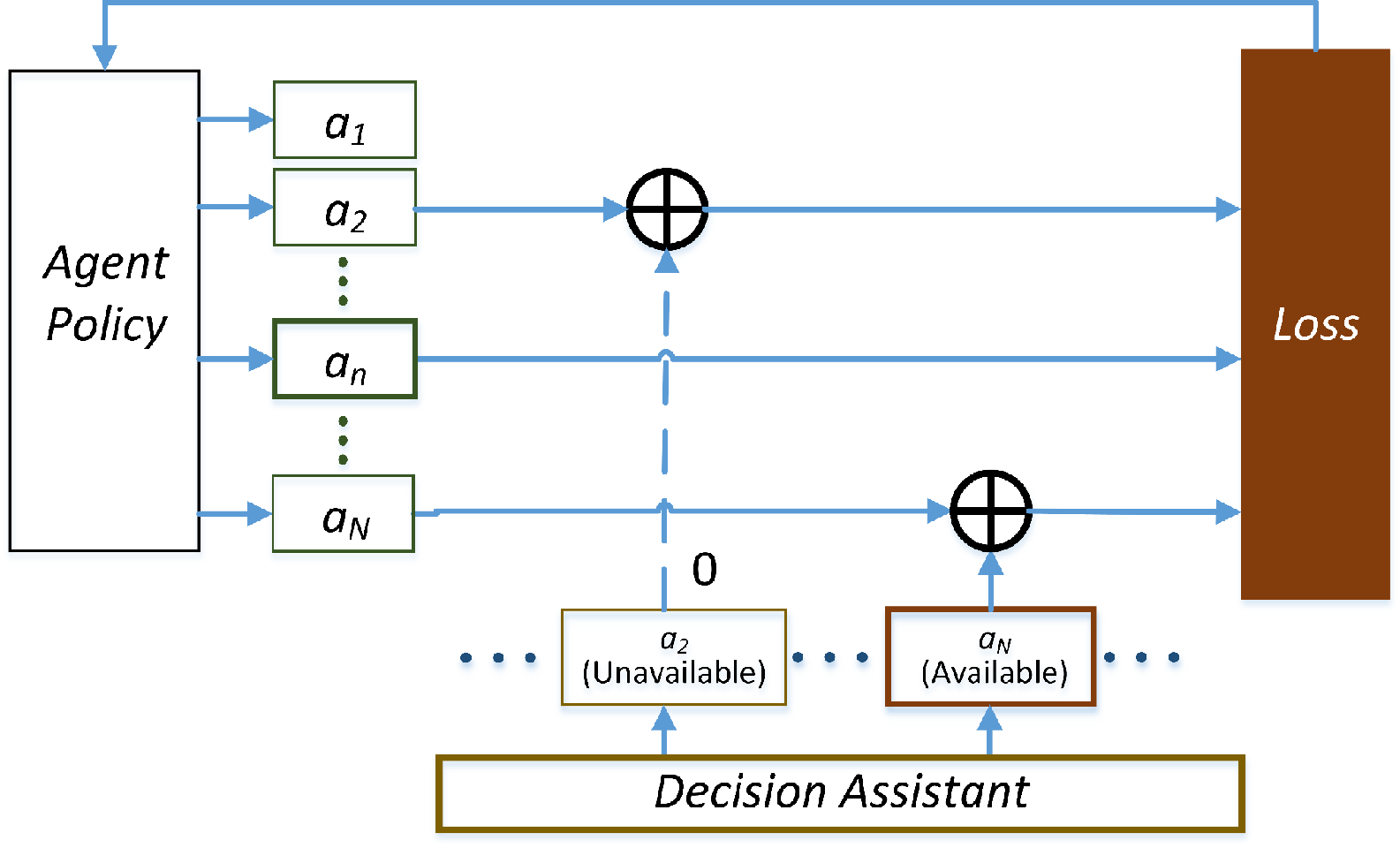}}
 \caption{The decision assistant in DRL.} \label{fig:DA}
\end{figure}

As shown in Fig. \ref{fig:DA}, each policy update deliberately includes a subset of currently infeasible actions identified by the feasibility rules. The policy is trained to output the baseline for those actions, ensuring they receive minimal preference and their gradients rapidly vanish. Because infeasible actions are generated deterministically by prior knowledge (task-completion status, coverage/link-budget checks under WPT scheduling, etc.) rather than by exploration, the range of effective exploration space is reduced, which improves the convergence efficiency of the DRL agent. The complete procedure of the proposed decision-assisted DSAC (DA-DSAC) is summarized in Algorithm \ref{Algorithm DSAC}.

\begin{algorithm}[t!]
\caption{\textbf{DA-DSAC-Based Optimization in UAV-Enabled WPT Network}: }\label{Algorithm DSAC}
\begin{algorithmic}[1]
 \makeatletter\setcounter{ALG@line}{0}\makeatother
 \State Initialize parameters in UAV-Enabled WPT networks and semantic transmission framework.
 \State Initialize DRL learning variables.
 \Repeat:
 \For {time slot 0, 1, ..., T}
 \State Choose an action $a(t)$ based on the current policy
 \Statex \phantom{0}\phantom{0}\phantom{0}\phantom{0}\phantom{0}\phantom{0}and state $f(t)$.
 \State Obtain the next state $f(t+1)$ and reward $r(t)$ from
 \Statex \phantom{0}\phantom{0}\phantom{0}\phantom{0}\phantom{0}\phantom{0}the environment.
 \State Generate the sample $\{f(t),a(t),r(t),f(t+1)$ for
 \Statex \phantom{0}\phantom{0}\phantom{0}\phantom{0}\phantom{0}\phantom{0}time slot $t$.
 \State Generate samples for unavailable action space.
 \State Save samples to the replay buffer.
 \State Update based on \eqref{eq:dsacA6}.
 \EndFor
 \For {each episode}
 \State Generate a minibatch including several samples
 \Statex \phantom{0}\phantom{0}\phantom{0}\phantom{0}\phantom{0}\phantom{0}from the replay buffer.
 \State Update DRL agent.
 \EndFor
 \Until Convergence.
\end{algorithmic}
\end{algorithm}

\section{Simulation Results} \label{sec:sim}
In simulations, we assume the number of EHBS $M = 10$, the number of GUs $N = 5$, EHBSs and GUs are randomly distributed within a 200 m × 200 m area, the UAV's initial altitude is 50 m, the maximum speed of UAV $v_{\rm \max} = 20$ m/s, the time slot duration is 1 s, the UAV altitude limitation $z_{\min} = 45$ m, $z_{\max}$ = 70m, the Rician factor $\kappa = 10$ dB, path loss exponents $\alpha_L = 2$ and $\alpha_N = 2.5$, the transmit power of EHBS is 1 W, the noise power at ground users is -10 dBm, the maximum harvested power for the UAV $W$ = 20 mW, the energy harvesting circuit physical hardware constraints $\xi$ = 1500, $\epsilon$ = 0.0022 \cite{7843670}, the bandwidth is 20 MHz, the size of raw data is 3.5 MB, the number of total time slots $T = 100$, the number of tasks $K = 50$, the latency threshold is randomly generated between 0.5 s and 2 s for tasks, the quality threshold is randomly generated between 10 and 25 (PSNR), The speed range for GUs' random waypoint model is (0, 5), the weight factors in the SEM are set as 0.5, the discount factor $\gamma = 0.99$. The ground users generate transmission tasks according to FTP Model 3. Moreover, we introduce the multi-agent proximal policy optimization (MAPPO) \cite{MPPO}, `Fixed $\beta$' ($\beta$ is fixed in the optimization) and `Hovering UAV' (The UAV remains hovering all the time) as the baselines in simulations.

\begin{figure}[t!]
  \centering
  \centerline{\includegraphics[scale=0.6]{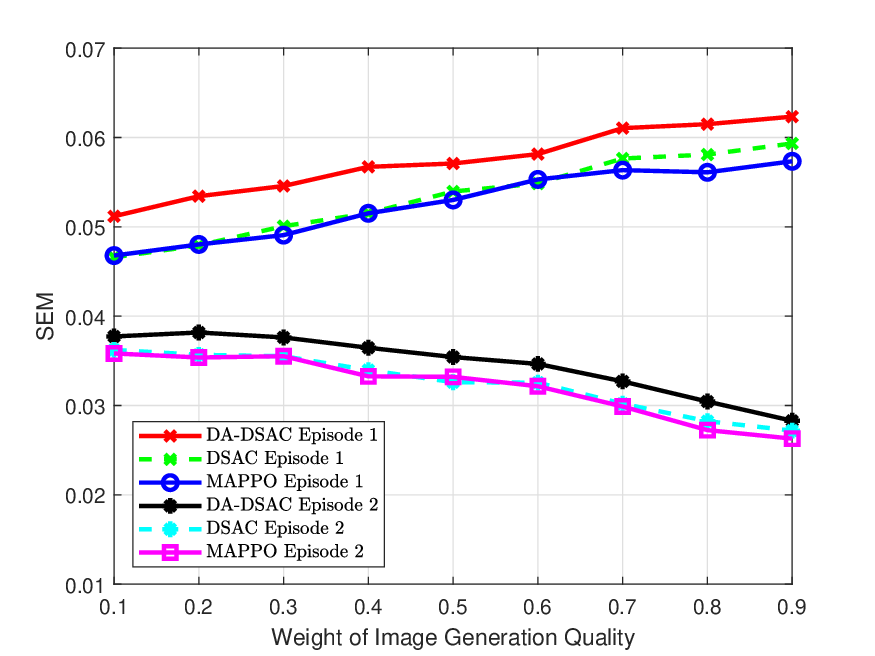}}
 \caption{\small Semantic efficiency vs the weight factor of generation quality.} \label{fig:R1}
\end{figure}
Fig. \ref{fig:R1} illustrates the changes of semantic efficiency as the weight factor of SEM changes. In this figure, we present the results of two different episodes. In Episode 1, as the image-quality weight increases, the SEM value also rises accordingly. This result indicates that in this episode, the importance of image quality in the generated tasks is more significant, while the delay constraint is relatively relaxed. However, in Episode 2, we can see an opposite trend. As the image-quality weight increases, the SEM decreases from approximately 0.04 to 0.03. This is because in Episode 2, the randomly generated tasks have tighter delay constraints, while the threshold for generation quality is relatively lower. Therefore, in practice, the weight factors can be determined according to application requirements or preset service policies. Moreover, it can be observed that in both two episodes, the proposed algorithm DA-DSAC outperforms MAPPO and DSAC, this result verifies the superiority of the proposed DRL method.

\begin{figure}[t!]
  \centering
  \centerline{\includegraphics[scale=0.6]{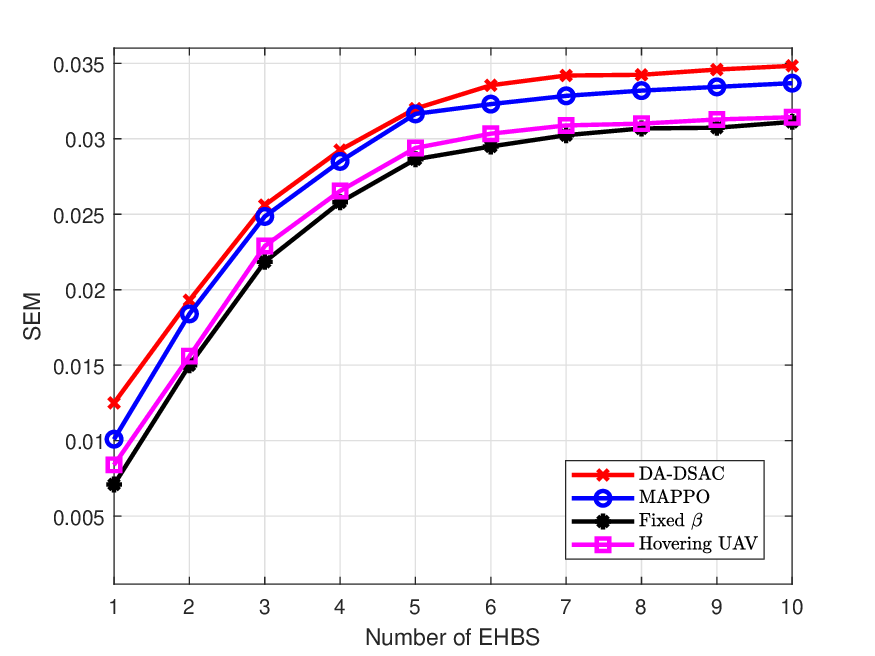}}
 \caption{\small Semantic efficiency vs the number of EHBS.} \label{fig:R2}
\end{figure}
Fig. \ref{fig:R2} shows the relationship between the number of EHBSs and the semantic communication efficiency. As the number of EHBSs increases, the SEM value initially improves significantly. When the number of EHBSs increases from 3 to 5, the SEM rises from 0.027 to 0.033. However, when the number of EHBSs is larger than 6, the growth of SEM becomes less noticeable. This is because each time an EHBS is selected, both the signal strength between the UAV and the EHBS, and that between the UAV and ground users, need to be considered simultaneously under the UAV’s position constraint. However, when the number of EHBSs becomes sufficiently large, the performance gain from further increasing EHBSs becomes marginal. Moreover, the baseline results in this figure demonstrate the advantages of the proposed algorithm, and highlight the importance of optimizing UAV trajectory and wireless power transfer time factor.

\begin{figure}[t!]
  \centering
  \centerline{\includegraphics[scale=0.6]{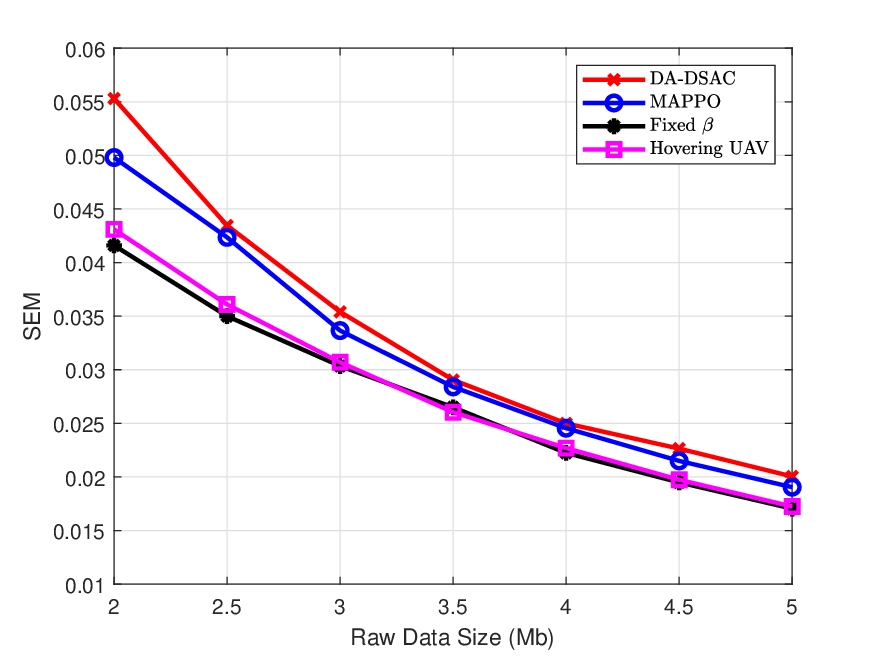}}
 \caption{\small Semantic efficiency vs raw data size.} \label{fig:R3}
\end{figure}
The impact of raw data size on semantic communication efficiency is shown in Fig. \ref{fig:R3}. It is clearly observed that as the raw data size increases, the SEM value significantly decreases. When the raw data size reaches 5 MB, the SEM drops to around 0.022. The reason for this result is, although semantic communication can significantly compress the raw data, obtaining a better compression ratio results in generation performance degradation. Conversely, larger transmission delays are introduced with less compression performance. Therefore, the increase in raw data size brings challenges to the efficiency of semantic communications.

\begin{figure}[t!]
  \centering
  \centerline{\includegraphics[scale=0.6]{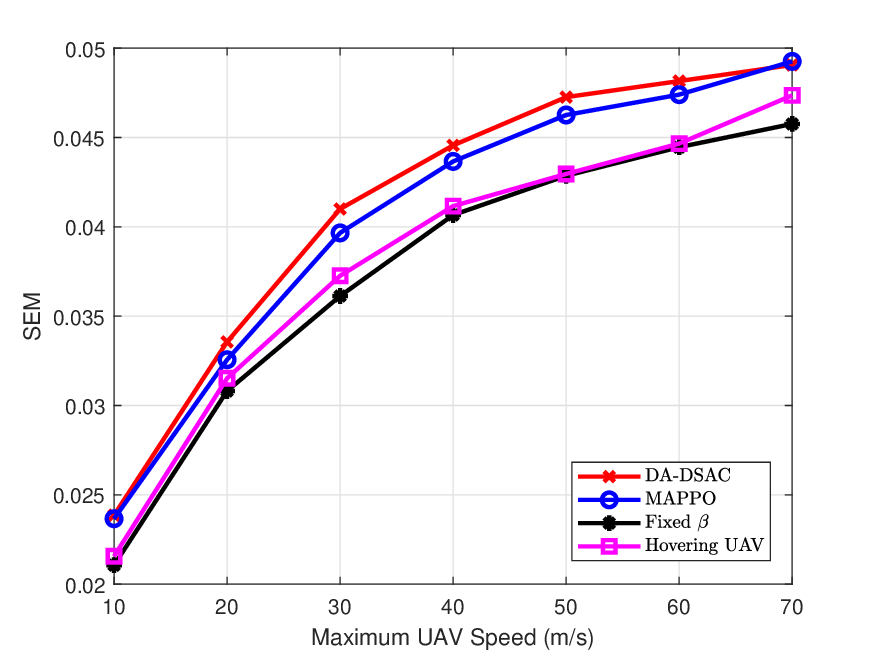}}
 \caption{\small Semantic efficiency vs maximum UAV speed.} \label{fig:R4}
\end{figure}
Fig. \ref{fig:R4} presents the impact of UAV mobility on system performance. As the UAV’s speed increases, the semantic communication efficiency also improves. When the UAV’s maximum speed reaches 60 m/s, the SEM value increases to nearly 0.048. This is because a higher maximum speed provides greater mobility ability to the UAV, so the UAV can quickly reach areas and find more favorable deployment locations that are difficult to access at lower speeds. On the other hand, the growth of SEM slows down when the maximum UAV speed approaches around 70 m/s, since such a high speed is already sufficient for UAV rapid deployment within the service area.

\begin{figure}[t!]
  \centering
  \centerline{\includegraphics[scale=0.6]{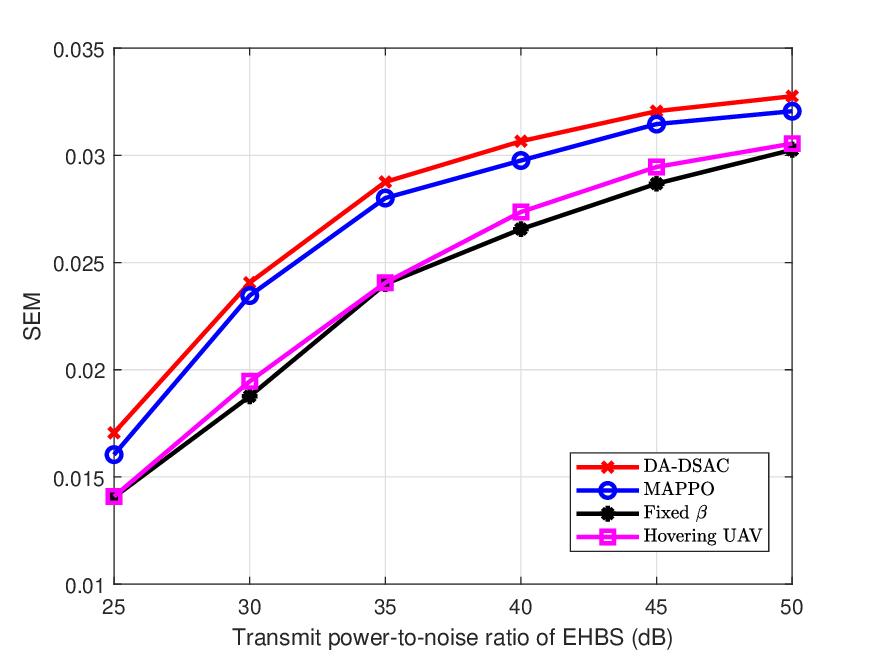}}
 \caption{\small Semantic efficiency vs the transmit power-to-noise ratio of EHBS.} \label{fig:R5}
\end{figure}
Furthermore, we investigate the impact of increasing EHBS transmit power on the overall semantic communication efficiency in Fig. \ref{fig:R5}. As we expected, the UAV can harvest more energy to increase its transmission power with higher EHBS power, or spend less time on energy harvesting and allocate more time for data transmission. Therefore, increasing the EHBS transmit power can directly enhance the semantic communication efficiency.

\begin{figure}[t!]
  \centering
  \centerline{\includegraphics[scale=0.6]{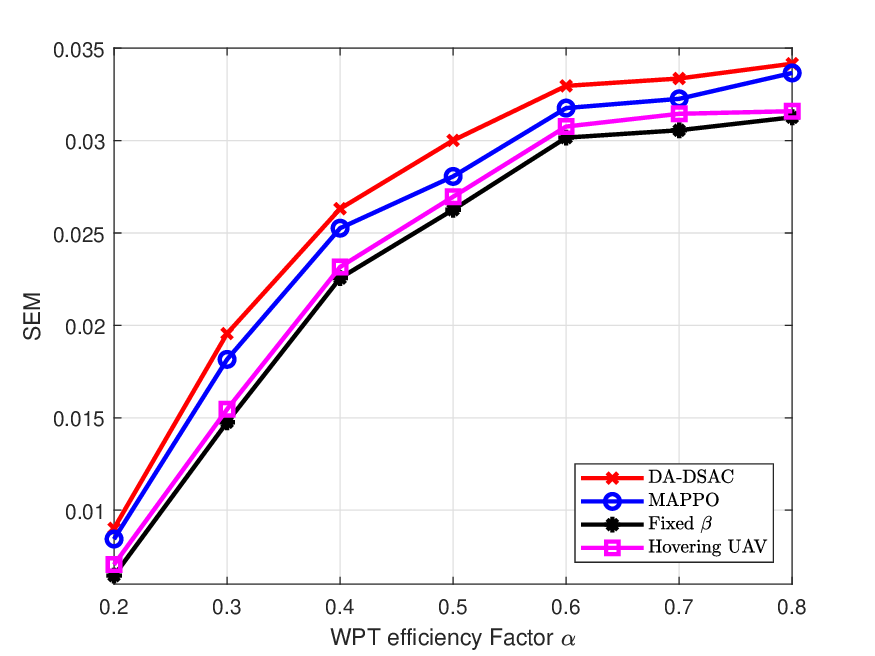}}
 \caption{\small Semantic efficiency vs WPT efficiency Factor.} \label{fig:R6}
\end{figure}
Fig. \ref{fig:R6} illustrates the impact of WPT efficiency on the semantic communication efficiency of the UAV. Similar to increasing the EHBS transmit power, improving the WPT transmission efficiency achieves a comparable effect by reducing the UAV’s energy harvesting time or allowing it to transmit with higher power, thereby further enhancing spectrum utilization. This demonstrates that the integration of WPT and semantic communication is highly suitable for future wireless communication networks, particularly for UAVs that are constrained by limited energy resources.

\begin{figure}[t!]
  \centering
  \centerline{\includegraphics[scale=0.6]{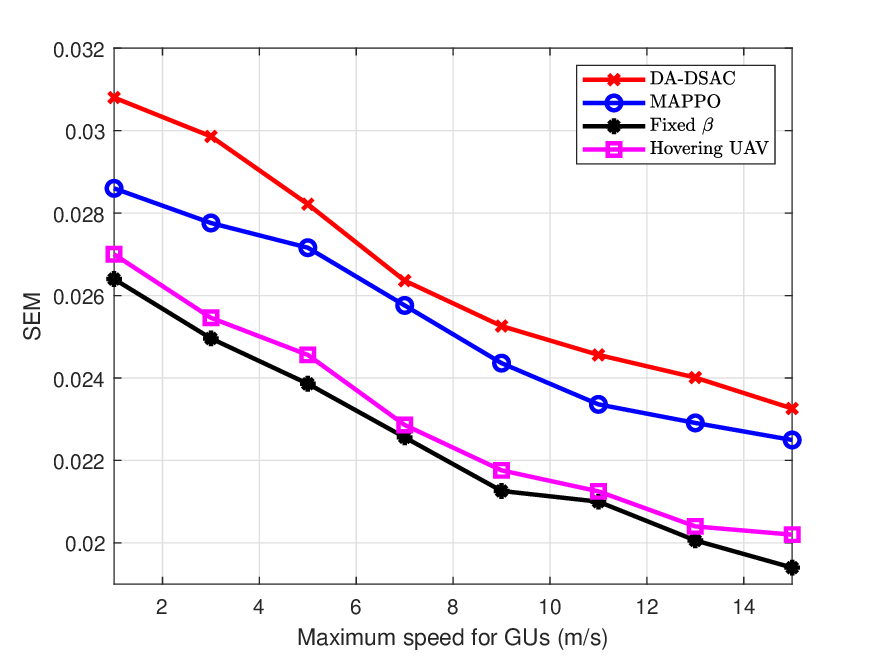}}
 \caption{\small Semantic efficiency vs maximum GU speed.} \label{fig:R7}
\end{figure}
To investigate the impact of ground user mobility on system communication performance, Fig. \ref{fig:R7} varies the maximum speed constraint in the ground user random waypoint model. When the maximum GU speed increases to 12 m/s, the semantic efficiency metric is decreased to around 0.025. This result shows that the ground user mobility indeed affects the semantic communication efficiency in the proposed network. As the ground users’ moving speed increases, the system’s dynamics and uncertainty also grow, so it is more difficult for the UAV to quickly find an optimal position to maximize the semantic communication efficiency. Therefore, mobility prediction of users will be an important research direction for future communication networks.

\section{Conclusion}\label{sec:con}
This paper investigates a hybrid bit and semantic communication framework in UAV-enabled WPT networks, where ground users can move randomly. We proposed a semantic communication efficiency metric to balance the tradeoff between image generation quality and transmission latency. To maximize this metric, we jointly optimized the UAV trajectory, EHBS selection, user association, semantic communication mode selection, and energy harvesting time factor. To address this complex long-term optimization problem, we developed a decision-assisted DRL algorithm to enhance the convergence performance of DRL. Simulation results verified the advantages of the proposed transmission framework and optimization algorithm, demonstrating their effectiveness in achieving adaptive and energy-efficient communication for future large-scale UAV networks and wireless systems. In future work, we plan to extend this study to multi-UAV and multi-source energy harvesting scenarios, and further explore the tradeoff between computation and communication in semantic communication systems.

\bibliographystyle{IEEEtran}
\bibliography{IEEEabrv,ref}

\begin{thebibliography}{10}
\providecommand{\url}[1]{#1}
\csname url@samestyle\endcsname
\providecommand{\newblock}{\relax}
\providecommand{\bibinfo}[2]{#2}
\providecommand{\BIBentrySTDinterwordspacing}{\spaceskip=0pt\relax}
\providecommand{\BIBentryALTinterwordstretchfactor}{4}
\providecommand{\BIBentryALTinterwordspacing}{\spaceskip=\fontdimen2\font plus
\BIBentryALTinterwordstretchfactor\fontdimen3\font minus
  \fontdimen4\font\relax}
\providecommand{\BIBforeignlanguage}[2]{{%
\expandafter\ifx\csname l@#1\endcsname\relax
\typeout{** WARNING: IEEEtran.bst: No hyphenation pattern has been}%
\typeout{** loaded for the language `#1'. Using the pattern for}%
\typeout{** the default language instead.}%
\else
\language=\csname l@#1\endcsname
\fi
#2}}
\providecommand{\BIBdecl}{\relax}
\BIBdecl

\bibitem{lu2024tutorial}
Z.~Lu, R.~Li, K.~Lu, X.~Chen, E.~Hossain, Z.~Zhao, and H.~Zhang,
  ``Semantics-empowered communications: A tutorial-cum-survey,'' \emph{IEEE
  Commun. Surveys Tuts.}, vol.~26, no.~1, pp. 41--79, 2024.

\bibitem{xie2021tsp}
H.~Xie, Z.~Qin, G.~Y. Li, and B.-H. Juang, ``Deep learning enabled semantic
  communication systems,'' \emph{IEEE Trans. Signal Processing}, vol.~69, pp.
  2663--2675, Apr. 2021.

\bibitem{bourtsoulatze2019deepjscc}
E.~Bourtsoulatze, D.~B. Kurka, and D.~G\"und\"uz, ``Deep joint source--channel
  coding for wireless image transmission,'' \emph{IEEE Trans. Cogn. Commun.
  Netw.}, vol.~5, no.~3, pp. 567--579, Sep. 2019.

\bibitem{gunduz2023jsac}
D.~G\"und\"uz, Z.~Qin, I.~E. Aguerri, H.~S. Dhillon, Z.~Yang, A.~Yener, K.-K.
  Wong, and C.-B. Chae, ``Beyond transmitting bits: Context, semantics, and
  task-oriented communications,'' \emph{IEEE J. Select. Areas Commun.},
  vol.~41, no.~1, pp. 5--41, Jan. 2023.

\bibitem{bi2015wpcn}
S.~Bi, Y.~Zeng, and R.~Zhang, ``Wireless powered communication networks: an
  overview,'' \emph{IEEE Wireless Communications}, vol.~23, no.~2, pp. 10--18,
  Apr. 2016.

\bibitem{11173626}
G.~Pan, Y.~Wu, Y.~Si, Z.~Hua, X.~Gao, S.~Wang, J.~An, and D.~Niyato,
  ``Distributed movable antenna systems for reliable, flexible, and resilient
  communications: Architectures, techniques, and challenges,'' \emph{IEEE
  Wireless Communications}, pp. 1--9, 2025, doi: 10.1109/MWC.2025.3599138.

\bibitem{zeng2016uav}
Y.~Zeng, R.~Zhang, and T.~J. Lim, ``Wireless communications with unmanned
  aerial vehicles: Opportunities and challenges,'' \emph{IEEE Commun. Mag.},
  vol.~54, no.~5, pp. 36--42, May. 2016.

\bibitem{9468714}
L.~Xie, X.~Cao, J.~Xu, and R.~Zhang, ``{UAV}-enabled wireless power transfer: A
  tutorial overview,'' \emph{IEEE Trans. Green Commun. Netw.}, vol.~5, no.~4,
  pp. 2042--2064, Dec. 2021.

\bibitem{yuan2021twc}
X.~Yuan, Y.~Hu, J.~Xu, and A.~Schmeink, ``Trajectory design for {UAV}-enabled
  multiuser wireless power transfer,'' \emph{IEEE Trans. Wireless Commun.},
  vol.~20, no.~1, pp. 26--42, Feb. 2021.

\bibitem{9763856}
L.~Yan, Z.~Qin, R.~Zhang, Y.~Li, and G.~Y. Li, ``Resource allocation for text
  semantic communications,'' \emph{IEEE Wireless Commun. Lett.}, vol.~11,
  no.~7, pp. 1394--1398, Jul. 2022.

\bibitem{9450827}
Z.~Weng and Z.~Qin, ``Semantic communication systems for speech transmission,''
  \emph{IEEE J. Select. Areas Commun.}, vol.~39, no.~8, pp. 2434--2444, Aug.
  2021.

\bibitem{9953076}
D.~Huang, F.~Gao, X.~Tao, Q.~Du, and J.~Lu, ``Toward semantic communications:
  Deep learning-based image semantic coding,'' \emph{IEEE J. Select. Areas
  Commun.}, vol.~41, no.~1, pp. 55--71, Jan. 2023.

\bibitem{10508293}
Z.~Ji, Z.~Qin, X.~Tao, and Z.~Han, ``Resource optimization for semantic-aware
  networks with task offloading,'' \emph{IEEE Trans. Wireless Commun.},
  vol.~23, no.~9, pp. 12\,284--12\,296, Sep. 2024.

\bibitem{witt2022}
K.~Yang, S.~Wang, J.~Dai, K.~Tan, K.~Niu, and P.~Zhang, ``Witt: A wireless
  image transmission transformer for semantic communications,'' \emph{IEEE
  International Conference on Acoustics, Speech and Signal Processing
  (ICASSP)}, Rhodes Island, Greece, Jun. 2023.

\bibitem{zhou2022adaptive}
Q.~Zhou, R.~Liu, J.~Zhang, R.~He, R.~Wang, and R.~Zhang, ``Adaptive bit rate
  control in semantic communication with incremental knowledge-based {HARQ},''
  \emph{IEEE Open J. Commun. Soc}, vol.~3, pp. 1075--1089, Jul. 2022.

\bibitem{11006980}
Z.~Zhao, Z.~Yang, M.~Chen, C.~Zhu, W.~Xu, Z.~Zhang, and K.~Huang,
  ``Energy-efficient probabilistic semantic communication over space-air-ground
  integrated networks,'' \emph{IEEE Trans. Wireless Commun.}, vol.~24, no.~10,
  pp. 8814--8829, Oct. 2025.

\bibitem{10854360}
X.~Peng, Z.~Qin, X.~Tao, J.~Lu, and K.~B. Letaief, ``A robust image semantic
  communication system with multi-scale vision transformer,'' \emph{IEEE J.
  Select. Areas Commun.}, vol.~43, no.~4, pp. 1278--1291, Apr. 2025.

\bibitem{9768113}
G.~Geraci, A.~Garcia-Rodriguez, M.~M. Azari, A.~Lozano, M.~Mezzavilla,
  S.~Chatzinotas, Y.~Chen, S.~Rangan, and M.~D. Renzo, ``What will the future
  of {UAV} cellular communications be? a flight from 5{G} to 6{G},'' \emph{IEEE
  Commun. Surveys Tuts.}, vol.~24, no.~3, pp. 1304--1335, Thirdquarter, 2022.

\bibitem{10770152}
H.~Chang, C.-X. Wang, R.~Feng, C.~Huang, L.~Hou, and E.-H.~M. Aggoune, ``Beam
  domain channel modeling and prediction for {UAV} communications,'' \emph{IEEE
  Trans. Wireless Commun.}, vol.~24, no.~2, pp. 969--983, Feb. 2025.

\bibitem{11216397}
X.~Xu, H.~Xu, D.~Wei, W.~Saad, M.~Bennis, and M.~Chen, ``Transformer-based
  collaborative reinforcement learning for fluid antenna system ({FAS})-enabled
  3{D} {UAV} positioning,'' \emph{IEEE Journal on Selected Areas in
  Communications}, vol.~44, pp. 1128--1143, 2026, doi:
  10.1109/JSAC.2025.3625523.

\bibitem{11123630}
Y.~Yao, W.~Xiao, P.~Miao, G.~Chen, H.~Yang, C.-B. Chae, and K.-K. Wong,
  ``{UAV}-relay-aided secure maritime networks coexisting with satellite
  networks: Robust beamforming and trajectory optimization,'' \emph{IEEE Trans.
  Wireless Commun.}, Early Access, 2025. DOI: 10.1109/TWC.2025.3596136.

\bibitem{10499205}
J.~Li, G.~Chen, T.~Zhang, W.~Feng, W.~Jiang, T.~Q.~S. Quek, and R.~Tafazolli,
  ``{UAV}-{RIS}-aided space-air-ground integrated network: Interference
  alignment design and {D}o{F} analysis,'' \emph{IEEE Trans. Wireless Commun.},
  vol.~23, no.~9, pp. 11\,678--11\,692, Sep. 2024.

\bibitem{10287142}
Y.~Liu, C.~Huang, G.~Chen, R.~Song, S.~Song, and P.~Xiao, ``Deep learning
  empowered trajectory and passive beamforming design in {UAV}-{RIS} enabled
  secure cognitive non-terrestrial networks,'' \emph{IEEE Wireless Commun.
  Lett.}, vol.~13, no.~1, pp. 188--192, Jan. 2024.

\bibitem{10483540}
Y.~Zhu, M.~Chen, S.~Wang, Y.~Hu, Y.~Liu, and C.~Yin, ``Collaborative
  reinforcement learning based unmanned aerial vehicle ({UAV}) trajectory
  design for 3{D} {UAV} tracking,'' \emph{IEEE Trans. Mobile Comput.}, vol.~23,
  no.~12, pp. 10\,787--10\,802, Dec. 2024.

\bibitem{10680056}
W.~Mao, Y.~Lu, G.~Pan, and B.~Ai, ``{UAV}-assisted communications in
  {SAGIN}-{ISAC}: Mobile user tracking and robust beamforming,'' \emph{IEEE
  Journal on Selected Areas in Communications}, vol.~43, no.~1, pp. 186--200,
  Jan. 2025.

\bibitem{11271691}
C.~Huang, G.~Chen, Z.~Xu, J.~Zhu, T.~Pan, R.~Tafazolli, and W.~Huang,
  ``Flexible reconfigurable intelligent surface-aided covert communications in
  {UAV} networks,'' \emph{IEEE Journal on Selected Areas in Communications},
  vol.~44, pp. 1577--1588, 2026, doi: 10.1109/JSAC.2025.3639197.

\bibitem{11107245}
L.~Pang, K.~Song, P.~Miao, C.~Huang, B.~Ji, Z.~An, and G.~Chen, ``Dynamic
  interference management by using the enhanced clustering and deep
  reinforcement learning in {VLC}-enabled {UAV} communication,'' \emph{IEEE
  Transactions on Consumer Electronics}, vol.~71, no.~3, pp. 7584--7596, Aug.
  2025.

\bibitem{11320813}
H.~Fan, Z.~Song, X.~Yang, T.~Li, S.~Wang, C.~Y. Leow, G.~Pan, and D.~Niyato,
  ``Joint optimization of delay and power efficiency of neighbor discovery in
  {UAV} networks,'' \emph{IEEE Transactions on Mobile Computing}, pp. 1--16,
  2025, doi: 10.1109/TMC.2025.3649859.

\bibitem{9769985}
K.~K. Nguyen, A.~Masaracchia, V.~Sharma, H.~V. Poor, and T.~Q. Duong,
  ``{RIS}-assisted {UAV} communications for {I}o{T} with wireless power
  transfer using deep reinforcement learning,'' \emph{IEEE J. Sel. Top. Signal
  Process.}, vol.~16, no.~5, pp. 1086--1096, Aug. 2022.

\bibitem{9708417}
H.~Ren, Z.~Zhang, Z.~Peng, L.~Li, and C.~Pan, ``Energy minimization in
  {RIS}-assisted {UAV}-enabled wireless power transfer systems,'' \emph{IEEE
  Internet Things J.}, vol.~10, no.~7, pp. 5794--5809, Apr. 2023.

\bibitem{10033084}
J.~Shi, P.~Cong, L.~Zhao, X.~Wang, S.~Wan, and M.~Guizani, ``A two-stage
  strategy for {UAV}-enabled wireless power transfer in unknown environments,''
  \emph{IEEE Trans. Mobile Comput.}, vol.~23, no.~2, pp. 1785--1802, Jan. 2024.

\bibitem{10399860}
H.~Yu, M.~Ju, and H.-C. Yang, ``Aggregate throughput maximization for
  {UAV}-enabled relay networks with wireless power transfer: Joint trajectory
  and power optimization,'' \emph{IEEE Trans. Veh. Technol.}, vol.~73, no.~6,
  pp. 8253--8265, Jun. 2024.

\bibitem{liew2022icassp}
Z.~Q. Liew, Y.~Cheng, W.~Y.~B. Lim, D.~Niyato, C.~Miao, and S.~Sun, ``Economics
  of semantic communication system in wireless powered {I}nternet of
  {T}hings,'' in \emph{IEEE International Conference on Acoustics, Speech and
  Signal Processing (ICASSP)}, Singapore, Singapore, May. 2022, pp. 8637-8641.

\bibitem{khalfet2025tcomm}
N.~Khalfet, C.~Psomas, S.~Chatzinotas, and I.~Krikidis, ``Semantic
  communications for simultaneous wireless information and power transfer,''
  \emph{IEEE Trans. Commun.}, vol.~73, no.~1, pp. 173--188, Jan. 2025.

\bibitem{delfani2025lcomm}
E.~Delfani and N.~Pappas, ``Semantics-aware updates from remote energy
  harvesting devices to interconnected {LEO} satellites,'' \emph{IEEE Commun.
  Lett.}, vol.~29, no.~8, pp. 1928--1932, Aug. 2025.

\bibitem{sang2025iotj}
N.~H. Sang, N.~D. Hai, N.~D.~D. Anh, N.~C. Luong, V.-D. Nguyen, S.~Gong,
  D.~Niyato, and D.~I. Kim, ``Wireless power transfer meets semantic
  communication for resource-constrained {I}o{T} networks: A joint transmission
  mode selection and resource management approach,'' \emph{IEEE Internet Things
  J.}, vol.~12, no.~1, pp. 556--569, Jan. 2025.

\bibitem{8698468}
C.~You and R.~Zhang, ``3{D} trajectory optimization in rician fading for
  {UAV}-enabled data harvesting,'' \emph{IEEE Trans. Wireless Commun.},
  vol.~18, no.~6, pp. 3192--3207, Jun. 2019.

\bibitem{7320989}
Y.~Liu, L.~Wang, S.~A. Raza~Zaidi, M.~Elkashlan, and T.~Q. Duong, ``Secure
  {D}2{D} communication in large-scale cognitive cellular networks: A wireless
  power transfer model,'' \emph{IEEE Trans. Commun.}, vol.~64, no.~1, pp.
  329--342, Jan. 2016.

\bibitem{7843670}
E.~Boshkovska, D.~W.~K. Ng, N.~Zlatanov, A.~Koelpin, and R.~Schober, ``Robust
  resource allocation for {MIMO} wireless powered communication networks based
  on a non-linear {EH} model,'' \emph{IEEE Transactions on Communications},
  vol.~65, no.~5, pp. 1984--1999, May. 2017.

\bibitem{vae}
D.~P. Kingma and M.~Welling, ``Auto-encoding variational bayes,''
  \emph{International Conference on Learning Representations (ICLR)}, Banff,
  Canada, Apr. 2014.

\bibitem{11207608}
C.~Huang, X.~Chen, G.~Chen, P.~Xiao, G.~Y. Li, and W.~Huang, ``Deep
  reinforcement learning-based resource allocation for hybrid bit and
  generative semantic communications in space-air-ground integrated networks,''
  \emph{IEEE J. Select. Areas Commun.}, Early Access, 2025. DOI:
  10.1109/JSAC.2025.3623157.

\bibitem{blip2}
J.~Li, D.~Li, S.~Savarese, and S.~Hoi, ``{BLIP}-2: Bootstrapping language-image
  pre-training with frozen image encoders and large language models,''
  \emph{International Conference on Machine Learning (ICML)}, vol. 202, Hawaii,
  USA, Jul. 2023, pp. 19730-19742.

\bibitem{clip}
A.~Radford, J.~W. Kim, C.~Hallacy, A.~Ramesh, G.~Goh, S.~Agarwal, G.~Sastry,
  A.~Askell, P.~Mishkin, J.~Clark, G.~Krueger, and I.~Sutskever, ``Learning
  transferable visual models from natural language supervision,''
  \emph{International Conference on Machine Learning (ICML)}, vol. 139, Jul.
  2021, pp. 8748-8763.

\bibitem{sdm}
R.~Rombach, A.~Blattmann, D.~Lorenz, P.~Esser, and B.~Ommer, ``High-resolution
  image synthesis with latent diffusion models,'' \emph{IEEE/CVF Conference on
  Computer Vision and Pattern Recognition (CVPR)}, New Orleans, USA, Jun. 2022,
  pp. 10684-10695.

\bibitem{vq-vae}
A.~van~den Oord, O.~Vinyals, and k.~kavukcuoglu, ``Neural discrete
  representation learning,'' \emph{Advances in Neural Information Processing
  Systems (NeurIPS)}, vol.~30, Long Beach, USA, Dec. 2017.

\bibitem{ho2021classifierfree}
J.~Ho and T.~Salimans, ``Classifier-free diffusion guidance,'' \emph{Neural
  Information Processing Systems (NeurIPS)}, Dec. 2021.

\bibitem{9448360}
J.~Duan, Y.~Guan, S.~E. Li, Y.~Ren, Q.~Sun, and B.~Cheng, ``Distributional soft
  actor-critic: Off-policy reinforcement learning for addressing value
  estimation errors,'' \emph{IEEE Trans. Neural Netw. Learn.}, vol.~33, no.~11,
  pp. 6584--6598, Nov. 2022.

\bibitem{11270936}
D.~Wei, X.~Xu, Y.~Liu, H.~Vincent~Poor, and M.~Chen, ``Optimizing model
  splitting and device task assignment for deceptive signal-assisted private
  multi-hop split learning,'' \emph{IEEE Journal on Selected Areas in
  Communications}, vol.~44, pp. 1512--1528, 2026, doi:
  10.1109/JSAC.2025.3637738.

\bibitem{10440193}
C.~Huang, G.~Chen, P.~Xiao, Y.~Xiao, Z.~Han, and J.~A. Chambers, ``Joint
  offloading and resource allocation for hybrid cloud and edge computing in
  {SAGINs}: A decision assisted hybrid action space deep reinforcement learning
  approach,'' \emph{IEEE J. Select. Areas Commun.}, vol.~42, no.~5, pp.
  1029--1043, May. 2024.

\bibitem{9321455}
C.~Huang, G.~Chen, and Y.~Gong, ``Delay-constrained buffer-aided relay
  selection in the {I}nternet of {T}hings with decision-assisted reinforcement
  learning,'' \emph{IEEE Internet Things J.}, vol.~8, no.~12, pp.
  10\,198--10\,208, Jun. 2021.

\bibitem{MPPO}
C.~Yu, A.~Velu, E.~Vinitsky, J.~Gao, Y.~Wang, A.~Bayen, and Y.~WU, ``The
  surprising effectiveness of {PPO} in cooperative multi-agent games,''
  \emph{Conference on Neural Information Processing Systems (NeurIPS)}, New
  Orleans, LA, Nov. 2022, pp. 24611-24624.

\end{thebibliography}
\end{document}